\documentclass[aps,prd,longbibliography,twocolumn,groupedaddress]{revtex4-2}

\usepackage{amsmath}
\usepackage{float}
\usepackage{graphicx}
\usepackage{color}
\usepackage[ruled,vlined]{algorithm2e}
\usepackage[thmmarks,amsmath,standard]{ntheorem}
\usepackage{physics}

\usepackage{layouts}

\usepackage{multirow}
\usepackage{xcolor}

\usepackage{hyperref}

\newcommand{\mc}[1]{\mathcal{#1}}

\newcommand{\genset}[1]{\langle#1\rangle}
\newcommand{\qwith}{\quad \mathrm{with} \quad}

\newcommand{\iden}{\vb{1}}
\newcommand{\defi}{:=}
\newcommand{\ifed}{=:}

\newcommand{\CNOT}{\mathrm{CX}}

\newcommand{\psth}{p^\star_{\mathrm{th}}}
\newcommand{\pth}{p_{\mathrm{th}}}

\DeclareMathOperator*{\SWAP}{SWAP}

\begin{document}

%Title of paper
\title{Lift-Connected Surface Codes}

\author{Josias Old}
\email[]{j.old@fz-juelich.de}

\author{Manuel Rispler}

\author{Markus Müller}

\affiliation{Institute for Quantum Information, RWTH Aachen University, Aachen, Germany}
\affiliation{Institute for Theoretical Nanoelectronics (PGI-2), Forschungszentrum Jülich, Jülich, Germany}

\date{\today}

\begin{abstract}
We use the recently introduced lifted product to construct a family of Quantum Low Density Parity Check Codes (QLDPC codes). The codes we obtain can be viewed as stacks of surface codes that are interconnected, leading to the name lift-connected surface (LCS) codes. LCS codes offer a wide range of parameters - a particularly striking feature is that they show interesting properties that are favorable compared to the standard surface code. For example, already at moderate numbers of physical qubits in the order of tens, LCS codes of equal size have lower logical error rate or similarly, require fewer qubits for a fixed target logical error rate. We present and analyze the construction and provide numerical simulation results for the logical error rate under code capacity and phenomenological noise. These results show that LCS codes attain thresholds that are comparable to corresponding (non-connected) copies of surface codes, while the logical error rate can be orders of magnitude lower, even for representatives with the same parameters.
This provides a code family showing the potential of modern product constructions at already small qubit numbers. Their amenability to 3D-local connectivity renders them particularly relevant for near-term implementations.
\end{abstract}

\maketitle

\newpage

\section{Introduction}
Quantum Error Correcting (QEC) codes are essential for the reliable operation of quantum computers~\cite{knill1998resilient}. 
Recent advances in hardware quality and qubit count of quantum computing devices enabled first experiments realizing different aspects of quantum error correction~\cite{ryan2021realization,postler2022demonstration,hilder2022fault,krinner2022realizing,acharya2023suppressing, Bluvstein2023}.

QEC codes encode logical qubits in a subspace of a higher dimensional Hilbert space. For stabilizer codes, the codespace is spanned by the simultaneous \(+1\)-eigenspace of a set of commuting operators, the stabilizer generators. 
The commonly shown parameter triple \([[n,k,d]]\) denotes the  number of physical qubits \(n\) employed by the error correcting code to encode \(k\) logical qubits with a minimum distance \(d\). The latter is defined as the minimum number of single qubit Pauli operators that have non-trivial action on the codespace, i.e.~the minimum weight of a logical operator. A code with distance \(d\) can correct at least for all errors up to weight \(t = \lfloor\frac{d}{2}\rfloor\).  
A \emph{code family} is specified by a sequence of stabilizer codes (that share most properties) with growing number of physical qubits. 
Promising QEC code families are \emph{surface codes}~\cite{kitaev1997quantum}. Surface codes only require nearest neighbor connectivity in a planar 2D architecture. This makes them especially suited for experimental platforms with manifest connectivity constraints, such as superconducting qubits. 
Additionally, surface codes have some of the highest known thresholds for realistic circuit level noise models~\cite{Stephens2014, fowler2012surface}. 

Despite these strong upsides, a major shortcoming of surface codes is the observation that a surface code patch essentially always encodes only a single logical qubit irrespective of the size of the patch. This is captured by the so-called code rate \(r = \frac{k}{n}\), i.e. the ratio of logical to physical qubit numbers, which is asymptotically zero for the surface code. In practical terms, this implies that scaling the code to improve its correction capabilities leads to a substantial qubit overhead. This in turn defines the challenge to find codes with better encoding rate while giving up as little as possible with respect to connectivity and logical (threshold) performance. To highlight a result in this direction, it has been shown that codes with bounded connectivity attaining a constant encoding rate could be used for constant overhead fault-tolerant quantum computation~\cite{gottesman2013fault}. 
Codes built from stabilizers with bounded degree of connectivity are called  \emph{quantum low-density parity check codes} (QLDPC).
In particular, a \((d_q,d_s)\)-QLDPC family of codes \(\mc{Q}_i\) with \(n_i \to \infty\) as \(i \to \infty\) has every qubit involved in a maximum of \(d_q\) stabilizer measurements and every stabilizer measurement involving at most \(d_s\) qubits, independent of \(i\)~\cite{mackay2004sparse}.  
A (Q)LDPC code family with both an asymptotically constant rate and a linear distance scaling, i.e.~\([[n, k \propto n, d \propto n]]\), is called \emph{"good"}. For a recent overview of QLDPC codes see~\cite{breuckmann2021quantum}.

The definition of QLDPC captures the bounded connectivity found in the surface (or closely related toric) and color codes. However despite technically being QLDPC, these are far from "good" both due to their vanishing rate and moreover their sub-optimal distance scaling $d\propto \sqrt{n}$. A big leap towards QLDPC codes with improved scalings was provided by the invention of the hypergraph product construction~\cite{tillich2013quantum}. Here, the product of two classical codes gives a quantum code, which remarkably preserves the (Q)LDPC property provided the two classical codes are LDPC to begin with. 
With this construction, the product of two good classical codes leads to a QLDPC code with constant rate while maintaining the distance scaling of the surface code ($d\propto \sqrt{n}$)~\cite{Sipser1996,tillich2013quantum}. After a series of breakthroughs, this line of research of product code constructions recently culminated in the remarkable discovery that  good quantum (LDPC) codes with constant rate and linear distance exist~\cite{breuckmann2021balanced, panteleev2022asymptotically}. Additional constructions of good QLDPC code families followed shortly after~\cite{leverrier2022quantum, dinur2022good}.
While this showcases "goodness" as an important guiding principle inspiring the search for better codes, a good but also practical code (family) with low enough number of qubits and embeddability so far remains elusive.

These results establish QLDPC codes as promising candidates for fault-tolerant quantum computation. By the same token, they define substantial challenges, both on the experimental and on the theory side. As a first challenge, known constructions of good codes involve prohibitively large numbers of qubits (\(\sim 10^6\)), leaving a substantial gap between what will be realistically available in near-term devices and what would be required for the above. 
As a second challenge, good QLDPC codes have been proven to require geometrically non-local connectivity. In fact, no-go theorems prevent good scaling of parameters if the connectivity of the code is restricted to some neighborhood that does not grow with the code~\cite{bravyi2009no,bravyi2010tradeoffs}.
Several proposals on circuit constructions and implementation of QLDPC codes with constrained connectivity in mind have been formulated~\cite{delfosse2021bounds, tremblay2022constant,bravyi2024high} and code constructions that explicitly leverage hardware capabilities like modularity are also considered~\cite{strikis2022quantum}.
Progress in platforms that allow for more connectivity opens the road to implement more advanced codes. Ion traps achieve all-to-all connectivity in single crystals mediated by motional modes, limited to a few tens of qubits~\cite{bruzewicz2019trapped}. This restriction can be overcome by the ability to shuttle ions~\cite{kaushal2020shuttling}. Shuttling also enables effective all-to-all connectivity in neutral atom arrays, where coherent control of hundreds of atoms can be realized~\cite{saffman2016quantum, moses2023race,bluvstein2022quantum, wu2022erasure, cong2022hardware, xu2023constant}. The potential for quantum error correction in neutral atom quantum processors has very recently been demonstrated in Ref.~\cite{Bluvstein2023}.

Further challenges from the conceptual side are logical gates and decoding. Decoders adapted from classical coding theory like Belief Propagation and Ordered Statistics Decoding (BP+OSD) perform reasonably well, but symmetry of certain quantum codes and large distances pose ongoing challenges~\cite{panteleev2021degenerate, roffe2020decoding, delfosse2022toward, berent2023software}.
While efficient decoders for good QLDPC codes are in principle available, the lack of codes with reasonable size prevents benchmarking these codes~\cite{leverrier2023efficient, gu2022efficient, dinur2022good}. 
Additionally, QLDPC codes with a property known as single-shotness allow for fault-tolerance with only a single round of (noisy) syndrome measurement, but at the cost of a large qubit overhead~\cite{quintavalle2021single, higgott2023improved}.

Fault-tolerant implementations of logical gates like transversal, i.e.~single qubit decomposable gates, in general rely on symmetries of the code~\cite{eastin2009restrictions, jochym2018disjointness, breuckmann2022fold}. Several approaches for general codes include teleportation based gates~\cite{brun2015teleportation}, generalized lattice surgery~\cite{cohen2022low} or codes specifically constructed to support certain gates~\cite{jochym2019fault}.  For hypergraph product codes, some implementations of gates have been proposed, but they remain short of generality or practicality~\cite{krishna2021fault, quintavalle2022partitioning}.

\subsection{Summary of Results}
In this work, we introduce a new QLDPC code family, which we call lift-connected surface (LCS) codes. For their construction, we employ the recently established lifted product. This technique is a key ingredient in the recent groundbreaking discovery of good QLDPC codes. Using comparatively simple input codes, we obtain QLDPC codes that can be straightforwardly seen as sparsely interconnected copies of surface codes, leading to the name LCS.
While their asymptotic scaling is not "good" in the strict sense of the term, i.e. in a constant rate regime, the distance grows proportional to the physical qubit number up to a maximum size (see discussion around Eq.~\ref{eqn:lcs_code_parameters}), they demonstrate the near-term potential of QLDPC (specifically lifted product) codes. 
We benchmark LCS codes under code capacity as well as phenomenological noise (i.e. noisy syndrome measurements) using an adapted BP+OSD decoder. We find that their asymptotic thresholds are comparable to disjoint copies of surface codes, summarized in Tab~\ref{tab:th}. However for concrete realizations they offer substantially lower logical error rates and higher pseudo-thresholds. We show that this carries over to circuit-level noise by constructing distance preserving circuits for small LCS codes. Using only a fraction of the number of physical qubits, they achieve the same logical error rates as copies of surface codes. Given that these advantages already appear for as few as tens of qubits, these results make LCS codes promising candidates for near-term QEC experiments.

\begin{table}
    \caption{Bit-flip noise thresholds of LCS codes compared to copies of surface codes, obtained from BP+OSD decoding of codes with increasing distance. For details of the families, refer to the text. All uncertainties here and in the following are readoff errors.}
    \label{tab:th}
    \centering
    \begin{ruledtabular}
    \begin{tabular}{p{3.3cm}cc}
        Code & \multicolumn{2}{c}{threshold} \\
        & code capacity  & phenomenological \\
        \hline
        LCS code family 1 & \(6.7 \pm 0.3 \%\) & \(2.9 \pm 0.1 \%\) \\
        LCS code family 2 & \(7.7 \pm 0.2\%\) & \(3.2 \pm 0.1 \%\) \\
        \(d\) copies of distance \(d\) surface codes & \(7.5  \pm 0.3 \%\) & \(2.9 \pm 0.1\%\) \\
        
    \end{tabular}
    \end{ruledtabular}
\end{table}

The manuscript is structured as follows. In Sec.~\ref{sec:LP}, we review the lifted product construction for quantum error correcting codes. 
In Sec.~\ref{sec:LCS}, we show how LCS codes are constructed from the lifted product and describe the code parameters and structure. In Sec.~\ref{sec:Noisy} we perform simulations over code capacity and phenomenological noise channels for several members of the LCS family, showing their error correction capabilities. Sec.~\ref{sec:ft} shows the construction and benchmarking of fault-tolerant syndrome readout circuits for representatives with small qubit numbers. Finally we outline a path towards logical gates in LCS codes in Sec.~\ref{sec:towards} before concluding in Sec.~\ref{sec:conclusion}.

\section{\label{sec:LP}The Lifted Product Construction}
The lifted product (LP) construction combines classical LDPC codes based on circulant permutation matrices with the hypergraph product (HGP) construction for quantum codes~\cite{panteleev2020quantum,tillich2013quantum}. 

\subsection{Hypergraph Product}
\begin{figure}
    \centering
    \includegraphics[width=\linewidth]{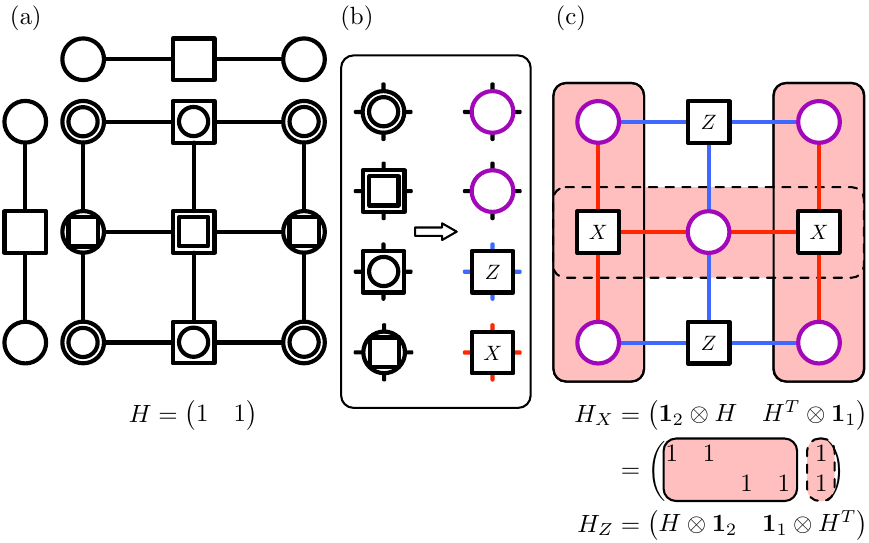}
    \caption{Pictorial representation of the hypergraph product (HGP) construction. (a) Shows the Cartesian product of two repetition codes of length two with parity check matrix \(H\). Black circles represent bits and black squares represent parity checks. With the prescription of (b), the Tanner graph (c) is obtained, which can be identified with the non-rotated \([[5,1,2]]\) surface code. Purple circles represent qubits, stabilizers are again black squares, where the edge color red (blue) denotes the Pauli-type $X$ ($Z$). We furthermore highlight the \(X\)-part as a guide to the reader to identify the blocks of the algebraic construction. The vertical stripes (solid border) correspond to the left \(\iden_2 \otimes H\) part of \(H_X\) as two copies of the original repetition code. The horizontal stripe (dashed border) comes from the right \(H^T \otimes \iden_1\) block, corresponding to (one copy of) the Tanner graph of the transposed repetition code.}
    \label{fig:hgp}
\end{figure}

A parity check matrix of a (classical code) $H$ can be represented by the so-called Tanner graph by identifying its rows (the parity checks $c$) with one type of node and its columns (the (qu)-bits $q$) with another type of node. An edge between nodes is drawn whenever the corresponding entry $H_{cq}$ is 1, making the graph manifestly bipartite~\cite{tanner1981a}.
Let \(\mc{T}_{\mc{C}_1}\) and \(\mc{T}_{\mc{C}_2}\) be the Tanner graphs of two classical codes \(\mc{C}_1\) and \(\mc{C}_2\) with binary parity check matrices \(H_1 \in \mathbb{F}_2^{m_1 \times n_1}, H_2 \in \mathbb{F}_2^{m_2 \times n_2}\) respectively. We will sometimes refer to these as \emph{base matrices}. 
The Tanner graph of the hypergraph product (quantum) code \(\mc{T}_{\mc{Q}}\) is based on the Cartesian product of the classical Tanner graphs \(\mc{T}_{\mc{C}_1}\) and \(\mc{T}_{\mc{C}_2}\). A graphical construction is shown in Fig.~\ref{fig:hgp}, for details refer to Ref.~\cite{tillich2013quantum}. Here, we review the algebraic construction rule when given two base matrices. The parity check matrices of the hypergraph product quantum CSS code are given by
\begin{align}
        H &= \mathrm{HGP}(H_1,H_2) = \mqty(0 & H_Z \\ H_X & 0) \qwith \\
        H_X &= \mqty(\iden_{n_1}  \otimes H_2 & H_1^T \otimes \iden_{m_2}) \\
        H_Z &= \mqty(H_1 \otimes \iden_{n_2} & \iden_{m_1}  \otimes H_2^T ) 
\end{align}
The CSS commutativity constraint \(H_Z H_X^T=0 \) is fulfilled by construction since
\begin{align}
        H_Z H_X^T &=  \mqty(H_1 \otimes \iden_{n_2} & \iden_{m_1}  \otimes H_2^T ) \mqty((\iden_{n_1}  \otimes H_2)^T \\ (H_1^T \otimes \iden_{m_2})^T) \nonumber\\
        &= (H_1 \otimes \iden_{n_2})(\iden_{n_1}  \otimes H_2^T ) \nonumber \\ 
        &\phantom{=} +(\iden_{m_1}  \otimes H_2^T ) (H_1 \otimes \iden_{m_2})\\
        &= H_1 \otimes H_2^T + H_1 \otimes H_2^T = 0. \label{eqn:hp_valid}
\end{align}

Any choice of binary matrices \(H_1, H_2\) gives a valid quantum code. Notably, surface codes can be obtained from taking the HGP of the \(\ell \times \ell+1\) parity check matrices of (classical) repetition codes, i.e.
        \begin{align}
        \label{eqn:pcm_rep}
        H^{(\ell)}_1 =H^{(\ell)}_2 = 
        \mqty(1      & 1      & 0      & \cdots & 0      & 0\\
              0      & 1      & 1      & 0      & \cdots & 0 \\
              \vdots &        &        &        &        & \vdots\\
              \vdots &        &        & \ddots & \ddots & 0\\
              0      & 0      & \cdots & 0 & 1      & 1 ) 
        \equiv H^{(\ell)}.
\end{align}
The resulting surface code \(\mathrm{HGP}(H^{(\ell)},H^{(\ell)})\) then has parameters \([[n,k,d]] =  [[(\ell+1)^2 + \ell^2, 1,\ell+1]]\).

If the base matrices are members of a \emph{good} classical LDPC code family with parameters \([n_{\mathrm{cl}}, k_{\mathrm{cl}} \propto n_{\mathrm{cl}}, d_{\mathrm{cl}} \propto n_{\mathrm{cl}}]\), then the resulting HGP codes have constant rate and distance \(d = \Omega(\sqrt{n})\)~\cite{tillich2013quantum}.

\subsection{Lifted Product}
To present the lifted product (LP) construction, we will follow the approach by Panteleev and Kalachev~\cite{panteleev2020quantum}. For a complementary approach see Ref.~\cite{breuckmann2021quantum}.
We restrict ourselves to a subset of LP codes, originally referred to as quasi-cyclic generalized hypergraph product codes~\cite{panteleev2021degenerate}. 
We first briefly review this generalization, before carefully explaining the implications for our construction in the next section.

A useful starting point before attempting to go beyond the HGP is to note a subtle requirement in fulfilling the CSS commutativity constraint of Eq.~\ref{eqn:hp_valid}.  Given the two parity check matrices $H_1$ and $H_2$ and using Dirac notation, it is indeed true that
\begin{align}
    &(H_1 \otimes \iden_{n_2} )( \iden_{n_1} \otimes H^T_2) = \\
    &= \sum_{abgh} H_{1,ab} H^T_{2,gh} \dyad{ag}{bh} \\
    &= H_1 \otimes H^T_2,
\end{align}
however when doing the analogous calculation 
\begin{align}
    (\iden_{m_1} \otimes H^T_2)(H_1 \otimes \iden_{m_2} ) &= \sum_{cgaf} H^T_{2,cg} H_{1,af} \dyad{ac}{fg},
\end{align}
the conclusion that this is also \(H_1 \otimes H^T_2\) rested on the assumption that all \(H^T_{2,cg}\) and \(H_{1,af} \) commute. While this commutativity is trivially true when the entries are numbers, it turns into a non-trivial requirement as soon as we want to promote the entries to higher-dimensional objects, e.g. matrices. In turn, this suggests that as long as we fulfill commutativity on this level, it will imply the fulfillment of the CSS constraint. One choice are elements of a commutative (matrix) ring, like circulant matrices of size \(L \times L\). A circulant \(C\) can be represented as sums of cyclic permutations \(P^{(i)}\),
\begin{align}
    C = \sum_{i = 0}^{L-1} c_i P^{(i)}
\end{align}
where \(c_i\) are binary coefficients and \(P^{(i)}\) denotes the \(i^{\mathrm{th}}\) cyclic (right) shift. We denote \(P^{(0)}\) by \(I\). For any circulant, we can give a binary representation such that \(\mathcal{B}_L(P^{(i)})\) is the \(i^{\mathrm{th}}\) cyclic (right) shift of the identity matrix \(\iden_L\). For example
\begin{align}
        \mathcal{B}_4(P^{(3)}) &= \mqty(0 & 0 & 0 & 1 \\ 1 & 0 & 0 & 0 \\  0 & 1 & 0 & 0 \\  0 & 0 & 1 & 0), \label{eqn:ex_lift_0}\\
        \mathcal{B}_3(I + P^{(1)}) &= \mqty(1 & 1 & 0 \\ 0 & 1 & 1 \\ 1 & 0 & 1), \label{eqn:ex_lift_1}\\
        \mathcal{B}_2\qty[\mqty(0 & I + P^{(1)}  \\ P^{(1)} & 0)] &= \mqty(0 & 0 & 1 & 1  \\ 0 & 0 & 1 & 1  \\ 0 & 1 & 0 & 0 \\ 1 & 0 & 0 & 0) . \label{eqn:ex_lift_2}
\end{align} 
The LP construction then takes the HGP of two matrices with circulants as entries and replaces these with the corresponding binary representations after having taken the product. This increases the number of qubits and parity checks by a factor of \(L\), which gives the procedure its name \emph{lifting}. 
Denoting matrices with circulant entries with a tilde, an LP code is obtained as 
\begin{align}
        H &= \mathrm{LP}(\tilde{H}_1,\tilde{H}_2) = \mathcal{B}_L(\tilde{H}) \qwith \\
        \tilde{H} &= \mqty(0 & \tilde{H}_Z \\ \tilde{H}_X & 0) \qwith \\
        \tilde{H}_X &= \mqty(\iden_{n_1}  \otimes \tilde{H}_2 & (\tilde{H}_1)^T \otimes \iden_{m_2}) \\
        \tilde{H}_Z &= \mqty(\tilde{H}_1 \otimes \iden_{n_2} & \iden_{m_1} \otimes (\tilde{H}_2)^T ).
\end{align}
Note that the transpose of a matrix with circulant entries \(\tilde{A} = (a_{ij})_{m \times n}\) is \(\tilde{A}^T = (a^T_{ji})_{n \times m}\) and it holds that \(\mathcal{B}_L(\tilde{A}^T) = \mathcal{B}_L(\tilde{A})^T\).

\begin{figure}
    \centering
    \includegraphics[width=\linewidth]{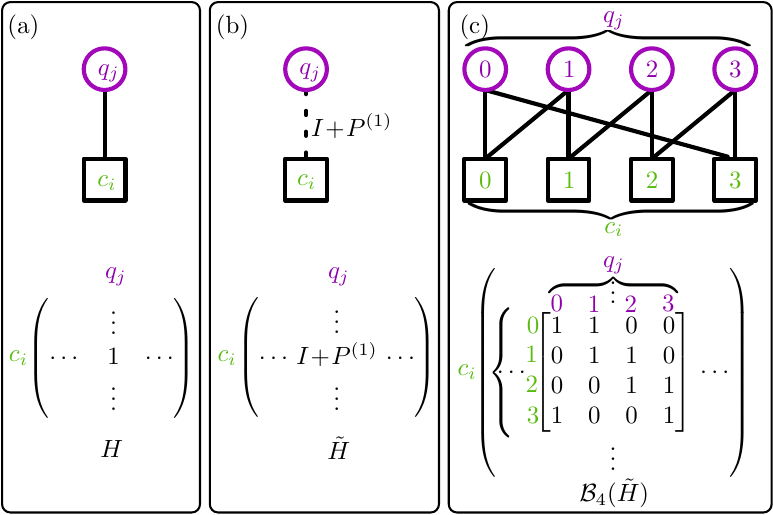}
    \caption{Lifting of edges of a graph that are part of the Tanner graph of a (quantum) code. \textbf{(a)} A nonzero entry of the parity check matrix \(H_{ij}\) corresponds to an edge between check \(c_i\) and qubit \(q_j\). \textbf{(b)} A circulant entry like \(I + P^{(1)}\) of a LP parity check matrix can be visualized by a label on the edge and a corresponding edge style. \textbf{(c)} The lift of the matrix corresponds to taking \(L\) (here \(L = 4\)) copies of the check and qubit nodes. The new connectivity is according to the binary representation of the circulant. In this example, the circulant \(I + P^{(1)}\) connects every copy of the check \((c_i, k)\) with two copies of qubits, \((q_j, k)\) and  \((q_j, k+1 \mod L)\).}
    \label{fig:lift}
\end{figure}

In graph-theoretical terminology, the parity check matrices \(H_X\) and \(H_Z\) are the \emph{biadjacency matrices} of the Tanner graphs. We show the lifting procedure for the example Eq.~\ref{eqn:ex_lift_1} in Fig.~\ref{fig:lift}. One edge of a Tanner graph of a HGP code with parity check matrix \(H\) between check \(c_i\) and qubit \(q_j\) is shown in Fig.~\ref{fig:lift}(a). In the LP construction, the entries of the final parity check matrix are replaced by circulants. This can be visualized by labelling the corresponding edge in the Tanner graph, as shown in Fig.~\ref{fig:lift}(b). The lift, in Fig.~\ref{fig:lift}(c) with lift parameter \(L=4\), translates to copying the check and qubit nodes \(L\) times and connecting them according to the non-zero entries of the circulant. In this example, the circulant \(I + P^{(1)}\) connects every copy of the check \((c_i, k)\) with two copies of qubits, \((q_j, k)\) and  \((q_j, k+1 \mod L)\).

\begin{figure*}
    \centering
    \includegraphics[width=0.7\linewidth]{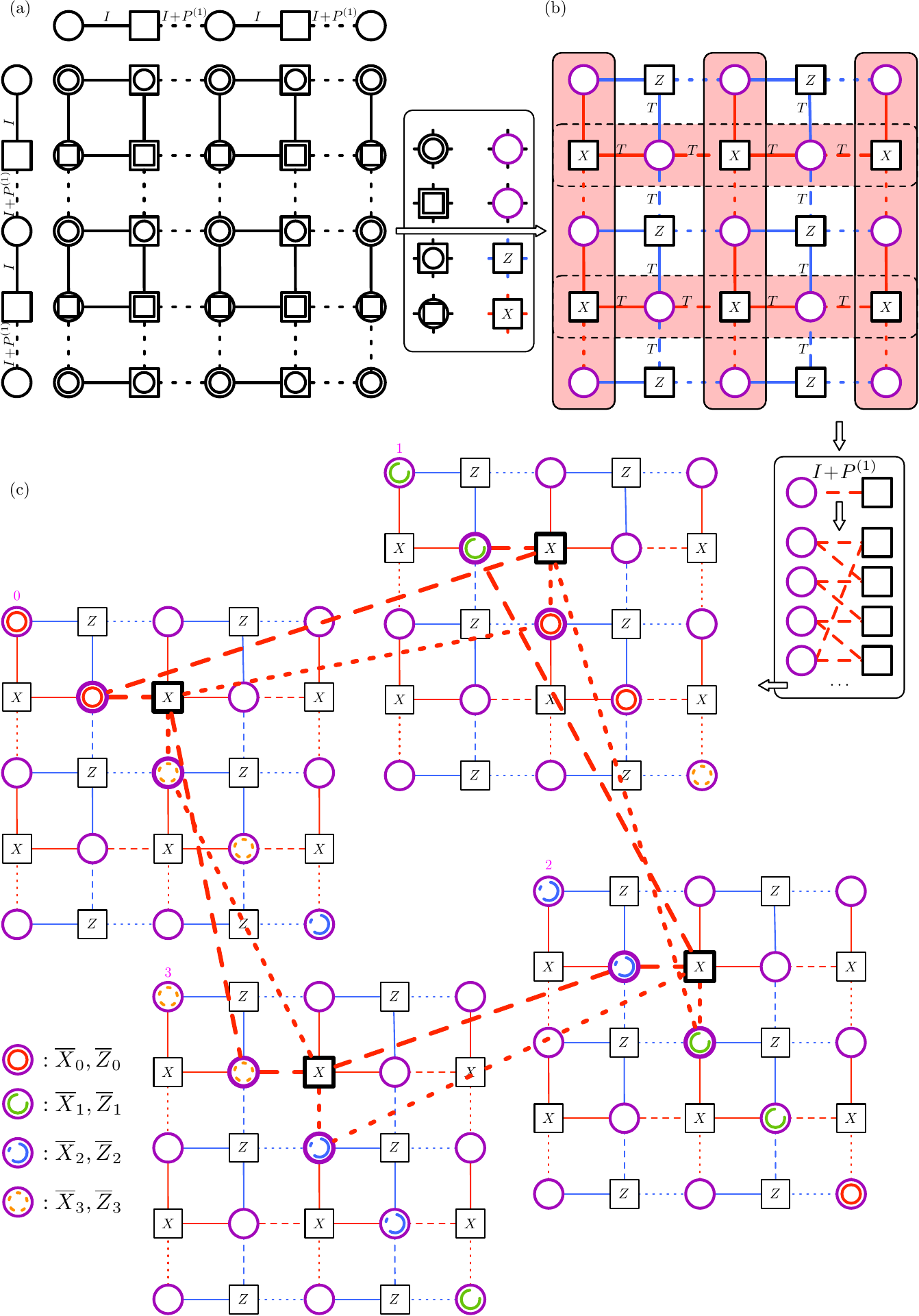}
     \caption{Pictorial representation of the lifted product construction for LCS codes with \((\ell,L) = (2,4)\). \textbf{(a)} First, we take the hypergraph product (HGP) of two repetition codes of length 3. The edges of the input Tanner graphs (top row and leftmost column) are now decorated and labeled to indicate that the scalar entries of the repetition code are being promoted (i.e. \emph{lifted}) to matrices $I$ (solid) and $I+P^{(1)}$ (dashed) respectively, which are both of dimension $L\times L$. This label carries through to the resulting Tanner graph. \textbf{(b)} Analogously to the HGP, vertices are respectively identified with qubits, \(X\)- and \(Z\)-checks. The edge label again indicates whether the corresponding entries are taken from the identity matrix $I$ (solid) or the circulant $I+P^{(1)}$ (dashed). To obtain the stabilizers from this picture, first of all observe that since all edges contain the $I$ matrix, we will obtain $L$ copies of the code that we would have obtained without lifting. The non-trivial extension comes from the edges containing the additional term $P^{(1)}$, which define the additional entries in the resulting stabilizer checks between the different copies. Also note that the edges of the transposed base graphs (indicated by $T$) also have to be lifted by the transposed circulant. This is exemplarily spelled out in \textbf{(c)}, where we show how the dashed edges of the given $X-$check in the LP are promoted to four $X-$checks on four copies of the underlying code and additional interconnections in the respective checks arise from the non-trivial circulants $P^{(1)}$ ($P^{(1)T}$). Note that the interconnections are highly structured, in particular they exclusively appear between neighboring code copies. We also indicate four logical representatives by small circles on the diagonal qubits of the surface code patches. }
    \label{fig:lp}
\end{figure*}

\section{The Lift-Connected Surface Codes \label{sec:LCS}}

\subsection{Construction and Parameters}
We can use the LP construction to algebraically build copies of the (non-rotated) surface code. To that end, we take base matrices of size \(\ell \times (\ell+1)\) of the same repetition code form as in Eq.~\ref{eqn:pcm_rep}. The binary entries are replaced the trivial circulant (\(P^{(0)} = I\)) of size \(L\) and zero circulant. The resulting matrix with circulant entries is denoted by \(\tilde{H}^{(\ell)}\).
Then the LP quantum code \(\mathrm{LP}_L(\tilde{H}^{(\ell)},\tilde{H}^{(\ell)})\) consists of \(L\) disjoint copies of distance \(d = \ell + 1\) surface codes. The parameters of the code are therefore
\begin{align}
        [[n,k,d]] &= [[((\ell+1)^2 +\ell^2)L , L, \ell + 1]], \nonumber \\
        (d_q,d_s) &= (4,4) .
\end{align}

This insight can be used to construct interconnected surface codes. Consider matrices of size  \(\ell \times \ell+1\) with
\begin{align}
        \tilde{H}^{(\ell)} &= \tilde{H}^{(\ell)}_{\text{rep.}} + \tilde{H}^{(\ell)}_{\text{int.}} 
        \\
        &= 
        \smqty(I      & I     & 0      & \cdots & 0      & 0\\
              0      & I      & I      & 0      & \cdots & 0 \\
              \vdots &        &        &        &        & \vdots\\
              \vdots &        &        & \ddots & \ddots & 0\\
              0      & 0      & \cdots & 0 & I      & I ) 
              +
        \smqty(0      & P^{(1)}     & 0      & \cdots & 0      & 0\\
              0      & 0      & P^{(1)}      & 0      & \cdots & 0 \\
              \vdots &        &        &        &        & \vdots\\
              \vdots &        &        & \ddots & \ddots & 0\\
              0      & 0      & \cdots & 0 & 0      & P^{(1)} ) 
              \\
        &= 
        \smqty(I      & I+P^{(1)}     & 0      & \cdots & 0      & 0\\
              0      & I      & I+P^{(1)}      & 0      & \cdots & 0 \\
              \vdots &        &        &        &        & \vdots\\
              \vdots &        &        & \ddots & \ddots & 0\\
              0      & 0      & \cdots & 0 & I      & I+P^{(1)} ).
\end{align}
The parity check matrices resulting from the first step (HGP) then naturally split into two parts,
\begin{align}
    \tilde{H}_X &= \smqty( \iden_{\ell + 1} \otimes \tilde{H}^{(\ell)}_{\text{rep.}} &\tilde{H}^{(\ell) T}_{\text{rep.}} \otimes  \iden_{\ell} ) + \smqty(   \iden_{\ell + 1} \otimes \tilde{H}^{(\ell)}_{\text{int.}}  & \tilde{H}^{(\ell) T}_{\text{int.}} \otimes \iden_{\ell}) \nonumber \\
     &\ifed \tilde{H}_{X,\text{surface}} +   \tilde{H}_{X,\text{interconnection}}  \\
    \tilde{H}_Z &= \smqty( \tilde{H}^{(\ell)}_{\text{rep.}}\otimes \iden_{\ell + 1} & \iden_{\ell} \otimes  \tilde{H}^{(\ell) T}_{\text{rep.}} ) + \smqty(  \tilde{H}^{(\ell)}_{\text{int.}} \otimes \iden_{\ell + 1} & \iden_{\ell} \otimes   \tilde{H}^{(\ell) T}_{\text{int.}}) \nonumber \\
    &\ifed \tilde{H}_{Z,\text{surface}} +   \tilde{H}_{Z,\text{interconnection}}.
\end{align}
The first addends \(\tilde{H}_{X,\text{surface}}, \tilde{H}_{Z,\text{surface}} \) have the same structure as surface codes. The lift therefore generates \(L\) copies of surface codes. The second addends \(\tilde{H}_{X,\text{interconnection}}, \tilde{H}_{Z,\text{interconnection}} \) act as additional connections in between the surface code patches. 
We therefore call the codes \(\mathrm{LP}_L(\tilde{H}^{(\ell)},\tilde{H}^{(\ell)})\) the \((\ell,L)\)-\emph{lift-connected surface codes} or simply \emph{LCS codes}.
The interconnections amount to at most two additional connections per check, because \(\tilde{H}^{(\ell)}_{\text{int.}}\) contains at most one entry per row and column.
The general form of the parity-check matrices is shown in App.~\ref{app:pcm}. The LCS code construction is shown pictorially in Fig.~\ref{fig:lp}.

The LCS codes are therefore  \((6,6)\)-QLDPC. The \emph{dimension} of a quantum CSS code is given by
\begin{align}
        k &= n - \rank(H_X) - \rank(H_Z).
\end{align}
Since both \(H_X\) and \(H_Z\) are already in a row echelon form and contain no zero rows, they are full rank and 
\begin{align}
        k = ((\ell+1)^2 +\ell^2)L - 2 \ell (\ell + 1) L = L.
\end{align}

\begin{figure*}
    \centering
    \includegraphics[width=\linewidth]{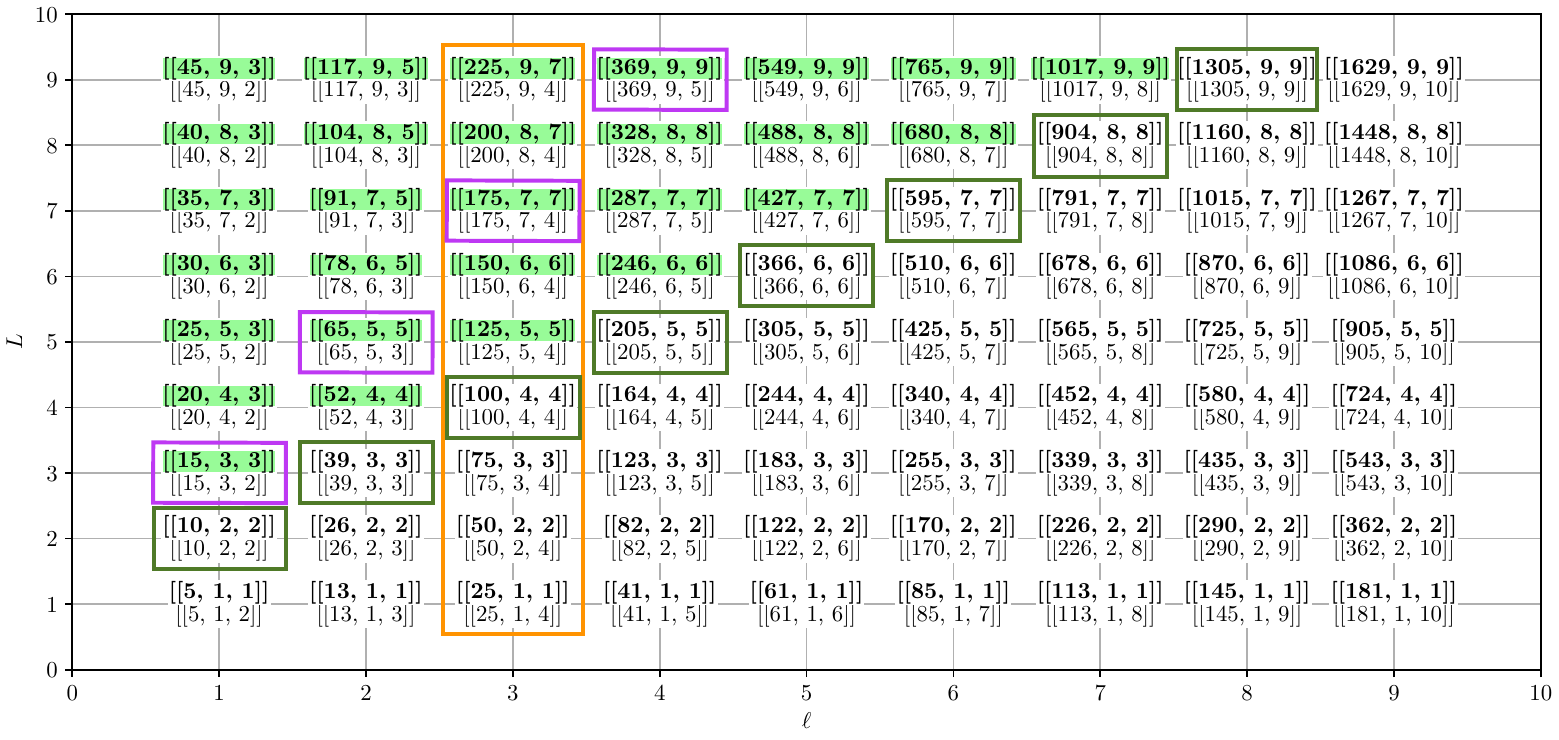}
    \caption{Code parameters \([[n,k,d]]\) of LCS (top, bold font) and copies of surface codes (bottom) for different \((\ell,L)\). The codes marked in dark green correspond to \(L = \ell + 1\) where both surface and LCS codes have the same parameters (family 1). Above that line for fixed \((\ell,L)\), the LCS codes have a larger distance compared to the copies of surface codes, indicated by the green shade behind the corresponding parameters. The purple boxes mark LCS code family 2 where for a fixed rate, codes with maximum distance and minimum number of physical qubits are chosen. The orange box (\(\ell = \mathrm{const.}\)) shows how for the LCS codes, \(k,d\) scale linearly in \(n\) up until \(L = 2 \ell+1\).}
    \label{fig:available_codes_lL}
\end{figure*}

There is no general recipe to get the distance of LP codes from the ingredients of the construction~\cite{panteleev2020quantum}. We can, however, calculate it via brute force by searching through all stabiliser equivalent representatives of logical operators or bound the minimum distance by probabilistic methods using the \texttt{GAP} package \texttt{QDistRnd}~\cite{pryadko2022qdistrnd}. We find for parameters \((\ell =1,L<100), (\ell=2,L<10), (\ell=3,L<5)\) that 
\begin{align}
    d = \min(L, 2 \ell + 1).
\end{align}
All data shown in this manuscript support this conjecture. In App.~\ref{app:pcm}, we show a constructive approach to establishing the minimum distance based on the block structure of the parity check matrices.
There, we also show sets of logical operators that can be understood from the point of view of interconnected surface codes. In (regular) surface codes, there exist representatives of $X$- and $Z$-logical operators with the same support on the diagonal of the surface code patch. These have weight $2 \ell + 1$ for surface code distance $\ell + 1$. While the canonical logical operators of surface codes do not "survive" the lift trivially, the operators on the diagonal can be lifted using consecutive shifts of the identity. In Fig.~\ref{fig:lp}, we indicate these logical operators, that have support on $2$, $2$ and $1$ qubits of successive copies of surface codes.
In summary, the parameters of the LCS codes are
\begin{align}
        [[n,k,d]] &= [[((\ell+1)^2 +\ell^2)L , L, \min(L, 2\ell + 1)]], \nonumber \\
        (d_q,d_s) &= (6,6) \label{eqn:lcs_code_parameters}.
\end{align}
Available codes are shown in Fig.~\ref{fig:available_codes_lL} and discussed in the following section. When unclear, we denote the construction parameters as a superscript \([[n,k,d]]^{(\ell,L)}\).
Note that for fixed \(\ell\) and varying lift-parameter \(L \leq 2 \ell + 1\), 
%this constitutes a good QLDPC code family in the sense that 
both the number of encoded logical qubits and the distance scale linearly in the number of physical qubits with constant rate \(r = (2 \ell^2 + 2 \ell + 1)^{-1}\). 

Asymptotically for $L \propto \ell$, the LCS codes achieve a scaling \([[n, \mc{O}(n^{\frac{1}{3}}), \mc{O}(n^{\frac{1}{3}})]]\). Compared to other 3D-local codes, this is better than 3D surface and toric codes with \([[n, \mc{O}(1),  \mc{O}(n^{\frac{1}{3}})]]\) (or \([[n, \mc{O}(n^{\frac{1}{4}}), \mc{O}(n^{\frac{1}{4}})]]\) when considering \(n^{\frac{1}{3}}\) copies of 3D surface/toric codes). Recent advances on 3D local codes from subdivisions or layer codes achieve the optimal scaling \([[n, \mc{O}(n^{\frac{1}{3}}), \mc{O}(n^{\frac{2}{3}})]]\)~\cite{bravyi2009no,bravyi2010tradeoffs,haah2021a,lin2023geometrically,williamson2023layer}. It is, however, not clear if and how small examples can be constructed and how they perform.

\subsection{Connectivity} \label{sec:conn}
\begin{figure*}
\centering
      \includegraphics[width=0.6\linewidth]{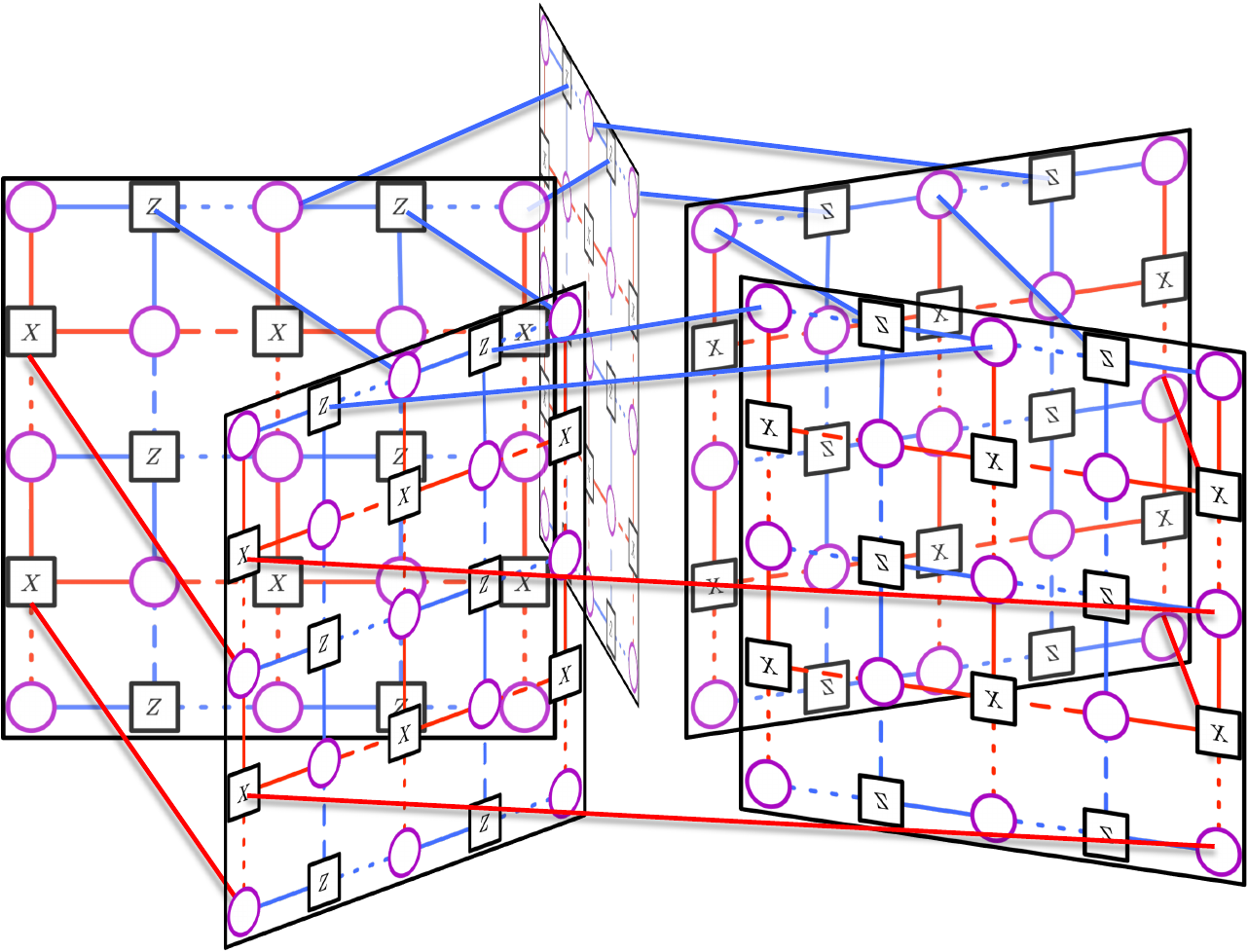}
      \caption{Possible arrangement of qubits of the \([[65,5,5]]^{(2,5)}\) LCS code, corresponding to \(5\) interconnected \([[13,1,3]]\) surface codes. While only a few interconnections of the patches are shown, note that all connectivity is restricted to neighbors of neighboring patches and therefore 3D-local.}
      \label{fig:3d_quasilocal}
\end{figure*}

Firstly, the LCS codes are QLDPC with \((d_q,d_s)_{\mathrm{lcs}} = (6,6)\) being slightly larger than the surface codes' \((d_q,d_s)_{\mathrm{s}} = (4,4)\).
The additional connectivity however is limited and indeed 3D-local. 
We label qubits with the tuple \((q_i,s)\) where qubit \(q_i, i \in \{0,1, \dots, (\ell+1)^2 + \ell^2 -1\}\) of patch \(s \in \{0,1, \dots L-1\}\) (and checks correspondingly with \((c_i,s)\)). An edge then is the tuple \(e = [(c_i,s),(q_j,s')]\).
The circulant \(P^{(1)}\) in the interconnection part of the parity check matrices interconnects only neighboring patches \(s\) and \(s+1 \mod L\). The transposed circulant \(P^{(1)T} = P^{(L-1)}\) turns the cyclic right shift by one into a cyclic left shift by one, keeping the restricted connectivity by only interconnecting  neighboring patches \(s\) and \(s-1 \mod L\).
Hence, there only exist edges \(e = [(c_i,s),(q_j,s')]\) with \(s' = s+1 \mod L\).
Also, for every edge \(e = [(c_i,s),(q_j,s')]\) with \(s \neq s'\) there exists an edge \(e = [(c_i,s),(q_j,s)]\), i.e.~interconnection only involve checks and qubits that would have been connected within one patch. 
An example for \((\ell,L) = (2,5)\) is shown in Fig.~\ref{fig:3d_quasilocal}, where the underlying surface codes are arranged as slices of a torus. Only a few interconnections are shown, but these have a bounded length.
These observations imply that these codes might be suitable, e.g., for implementation in static 3D Rydberg atom array structures \cite{barredo2018synthetic}, without the need for shuttling. This only holds if the qubits at the outer edge are close enough to the respective ancillary qubit of the neighboring slice to perform entangling gates for stabilizer measurements. Concrete implementations and optimizations of qubit positions in 3D space are left as future work.

Note that the choice of taking \(P^{(1)}\) on every entry of \(\tilde{H}^{(\ell)}_{\text{int.}}\) is motivated by the restricted connectivity. Taking \(P^{(j)}\) with \(j \neq 1\) and \(j \neq \frac{L}{2}\) constructs equivalent codes up to permuting the surface code patches accordingly. For \(j = \frac{L}{2}\), \(P^{(j)T} = P^{(j)} \) and both additional connections of a check go to the same copy of the underlying surface code. In that case, the code decouples into \(\frac{L}{2}\) disjoint copies and has distance $d = 2$.
Putting different \(P^{(j)}\) matrices as entries of \(\tilde{H}^{(\ell)}_{\text{int.}}\) renders the codes non-local, which makes it less straightforward to analyze. However, heuristically we find that such codes typically have the same parameters compared to their 3D local partners with the same \((\ell,L)\).

\section{Code Performance over noisy quantum channels \label{sec:Noisy}}
\subsection{Sampling codes in quantum channels}
Here we discuss several standard methods for (Monte Carlo) sampling of the combination of a given decoder and a noise model. 
We consider two types of noise models:
\begin{itemize}
    \item \emph{Code capacity} model: Only the data qubits are affected by a single qubit i.i.d. noisy quantum channel (e.g.~a bit-flip channel). The syndrome measurement is assumed to be perfect, see also Alg.~\ref{alg:cc_sampling}.
    \item \emph{Phenomenological noise} model: This noise model extends the code capacity model by additionally including noisy syndrome measurements. This can e.g. be modeled by a perfect measurement followed by a  (classical) bit-flip channel on the result, see Alg.~\ref{alg:phn_sampling}.
\end{itemize}
Typically, to ensure a fault-tolerance level of \(t = \lfloor\frac{d}{2}\rfloor\), the number of noisy syndrome measurement cycles is chosen to be \(n_{\mathrm{sc}} = d\). This is based on the fact that, even if the \(t\) errors occur on measurements of the same syndrome in the measurements cycles, the \(d = 2t+1\) repetitions allow to correctly identify these errors. We can also think of this as defining repetition codes on the syndrome measurement outcomes, with extra variables introduced to model syndrome flips. In order to preserve the distance \(d\) of the code, the length of these repetition codes is chosen to be \(d\). This procedure is also shown in Fig.~\ref{fig:rep_code_tg_phenomenological} for a three-bit repetition code.

\subsection{Decoding}
A decoder takes syndrome data and potentially parameters of the noise model as input and returns an inference of the underlying error configuration (which in turn determines the appropriate recovery operation), that is consistent with the observed syndrome. Based on this error guess, a correction is applied that puts the state back to the codespace. The ideal maximum likelihood decoder picks a guess from the most likely error class, taking into account the code degeneracy, i.e. that distinct error configurations can be logically equivalent. Because this is in general computationally hard, practical decoders return an approximation, trading computational efficiency for non-optimality of the suggested recovery operation. E.g. the \emph{Most Likely Error} (MLE) decoder tries to determine the most likely error configuration, ignoring potential degeneracies. 
To be efficient, practical decoders generally try to exploit structure in the code and noise model, such that not every existing decoder is suitable for application to a general code. Notably, a matching-based decoder like \emph{minimum weight perfect matching} works well whenever elementary errors violate two parity checks~\cite{fowler2012towards}.  
This is not the case for LCS codes, since here a single error on a qubit violates up to \(3\) parity checks (considering only \(X\)- and \(Z\)- errors).
We therefore resort to two general purpose decoders, namely a \emph{MLE} decoder and a decoder based on \emph{Belief Propagation} and \emph{Ordered Statistics Decoding} (BP+OSD) \cite{poulin2008iterative,panteleev2021degenerate}.

Given our codes are symmetric with respect to \(X\)- and \(Z\), we consider one Pauli type only, e.g. (single qubit, i.i.d) bit-flips with probability \(p\), for a performance benchmark of the codes. The probability of a fixed configuration \(E = \{E_q\}_{q=0}^{n-1}\) of errors given an observed syndrome \(\vb{s}\) is
\begin{align}
    p(E|\vb{s}) &\propto \prod_{q = 0}^{n - 1} p(E_q) \prod_{c = 0}^{n_c - 1} \delta(\genset{E,S_c} = s_c) \\
    &= p^{w} (1-p)^{n-w} \\
    &\propto \qty(\frac{p}{1-p})^w,
\end{align}
where \(w\) is the weight of the configuration \(w = \abs{E}\). \(S_c \in \mc{S}\) denotes the stabilizer \(c\) and \(s_c\) the bit \(c\) of the observed syndrome \(\vb{s}\). We write \(\genset{P, P'}\) with \(P, P' \in \mc{P}_n\) for the function that indicates commutation,
\begin{align}
    \genset{P, P'} = \begin{cases}
        0 \qif \comm{P}{P'} = 0 \\
        1 \qif \acomm{P}{P'} = 0.
    \end{cases}
\end{align}
With that notation, the syndrome of error \(E\) can be written as
\begin{align}
    \sigma(E) = (\genset{E, S_c})_{c = 0}^{n_c - 1}.
\end{align}

\subsubsection{MLE Decoding with i.i.d. noise}
\label{subsec:MLE}
 For \(p < 0.5\) (\(p > 0.5\)), the most likely error (MLE) is the one with the lowest (highest) weight \(w\).
Landahl et al. give an intuitive implementation of an MLE decoder for quantum codes which we follow here~\cite{landahl2011fault}.
\paragraph{Code capacity noise}
Considering pure bit-flip noise, we have one (error free) syndrome \(\vb{s}_Z = \{s_{Z,c}\}_{c=0}^{n_c-1}\). Let \(\vb{x} = \{x_q\}_{q=0}^{n-1}\) be the binary representation of the Pauli error. Then MLE decoding can be stated as the optimization problem
\begin{align}
    \min &\sum_q x_q \\
    \text{subject to } &\bigoplus_{q \in \Gamma(c)} x_q = s_{Z,c} \qquad\forall c \\
    \qwith &x_q \in \mathrm{GF}(2) = \{0,1\}.
\end{align}
Here, \(\Gamma(c)\) is the set of qubits involved in stabilizer measurement \(c\) and \(\oplus\) indicates addition modulo \(2\). In words, this procedure states: minimize the weight of an error that is compatible with the syndrome. 
In vectorised form with \(Z\)-parity check matrix \(H_Z\), this can be written as 
\begin{align}
    \min \quad &\vb{1}^T \vb{x} \\
    \text{subject to } &H_Z \vb{x} = \vb{s}\!\mod 2 \\
    \qwith &\vb{x} \in \mathrm{GF}(2)^n = \{0,1\}^n.
\end{align}

\paragraph{Phenomenological noise}
With multiple rounds of noisy syndrome measurements, data errors after time step \(t\) are indicated by differences of syndrome bits from \(t\) to \(t+1\).
We write 
\begin{align}
    \Delta \vb{s}_t = \vb{s}_t - \vb{s}_{t-1} = \vb{s}_t + \vb{s}_{t-1} \mod 2 \quad \forall t
\end{align}
with \(\vb{s}_{t=0} \! \defi \! \vb{0}\) for the observed syndrome measurement outcomes. We introduce \(d\) data-error vectors \(\vb{x}_0, \dots \vb{x}_{d-1} \in \mathrm{GF}(2)^n\) and \(d\) syndrome measurement-error vectors \(\tilde{\vb{s}}_0, \dots \tilde{\vb{s}}_{d-1} \in \mathrm{GF}(2)^{n_c}\) for \(d\) noisy rounds of error correction. This allows one to formulate the MLE decoding optimization problem as
\begin{align}
    \min &\sum_t \vb{1}^T \vb{x}_t \\
    \text{subject to } &H \vb{x}_t + \tilde{\vb{s}}_t + \tilde{\vb{s}}_{t-1} = \Delta\vb{s}_t\!\mod 2 \quad\forall t,
\end{align}
requiring that the noise-less syndrome together with syndrome errors are compatible with the observed syndrome differences.
A last round of perfect syndrome measurements allows us to verify the correction returned by the optimization (see App.~\ref{alg:phn_sampling}). Vectorized, this reads
\begin{align}
    \min \quad &\vb{1}^T \vb{y} \\
    \text{subject to } &A \vb{y} = \Delta\vb{s} \mod 2 \label{eqn:phnsto}\\
    \qwith \vb{y} = (\vb{x}_0^T, \dots, \vb{x}_{d-1}^T, &\tilde{\vb{s}}_0^T, \dots \tilde{\vb{s}}_{d-1}^T) \in \mathrm{GF}(2)^{2 dn}.
\end{align}
Here 
\begin{align}
    A =
    \left(
    \begin{array}{c|c}
        \mqty{\dmat{H,H,\ddots,H}}  &  \mqty{\dmat{I\\I&I, \ddots, I&I}}.
    \end{array}
    \right)
\end{align}
We implement the MLE decoder in python using the \texttt{optlang} interface~\cite{jensen2017optlang} to the GNU Linear Programming Kit \texttt{GLPK} \cite{makhorin2012glpk} to solve the constrained minimization. The run-time of decoding is exponential in the number of qubits and (noisy) error correction rounds. In practice, this results in a limitation on the feasibility of simulating the decoding of codes. We restrict the simulations to codes with \(nd \leq 250\). 

\subsubsection{Belief Propagation + Ordered Statistics Decoding}
To enable benchmarking beyond low qubit numbers, we employ Belief Propagation and Ordered Statistics Decoding (BP+OSD). 
Belief Propagation (BP) is a well known decoder for classical codes, where it works particularly well for good LDPC expander codes~\cite{gallager1962low, mackay1997near}. BP calculates single-qubit error probabilities using statistical inference on the \emph{Tanner graph} of the code. While BP can be shown to be exact on trees, Tanner graphs usually contain cycles, which can hinder the performance of BP 
(see e.g. Refs.~\cite{poulin2008iterative, roffe2019quantum}).
One way to overcome these problems is to use Ordered Statistics Decoding (OSD) as a post-processor. While BP may be inconclusive in its output, it usually points to a subset of likely erroneous qubits, OSD then brute-forces the solution of the decoding problem on that subset~\cite{panteleev2021degenerate}.

We denote the qubits participating in syndrome measurement \(c\) by \(\Gamma(c) \subset{Q}\) with \(Q\) the set of all qubits. The syndrome measurements in which qubit \(q\) participates are \(\Gamma(q)\). 
To approximate single qubit marginal probabilities, two types of quantities, termed \emph{messages}, are updated iteratively until convergence or a maximum number of iterations is reached.  
The messages \(m_{q \to c}(E_q)\) from qubit \(q\) to check \(c\) correspond to the current estimate of error probabilities for \(E_q\) on qubit \(q\) given the estimates of all neighboring checks except \(c\),
\begin{align}
    \displaystyle m^{(i+1)}_{q \to c} &\propto p_0(E_q) \prod_{c' \in \Gamma(q) \setminus c} m^{(i)}_{c' \to q}(E_q) \label{eqn:bpq2c}.
\end{align}
The messages \(m_{c \to q}(E_q)\) from check \(c\) to qubit \(q\) collect incoming probability estimates of qubits except \(q\) and sum over all compatible configurations fixing the error \(E_q\),
\begin{align}
    m^{(i)}_{c \to q} &\propto \sum_{E_{\Gamma(c) \setminus q}} \delta[\sigma(E)_c = s_c] \prod_{q' \in \Gamma(c) \setminus q} m^{(i)}_{q' \to c}(E_{q'}). \label{eqn:bpc2q}
\end{align}
The product of all these messages incoming at a qubit node gives an estimate for the marginal probabilities of errors on that qubit, called \emph{belief},
\begin{align}
    b_{q}(E_q) &\propto p_0(E_q) \prod_{c \in \Gamma(q)} m^{(i_{\text{max}})}_{c \to q}(E_q).
\end{align}
The log-likelihood ratios 
\begin{align}
    l_q &= \log \frac{1-b_q(E)}{b_q(E)}
\end{align}
of qubit \(q\) are calculated and used to sort the qubits by their likelihood of being erroneous. In first order OSD, the parity check matrix \(H\) is truncated to a full rank square matrix \(H_{[R]}\), with \(R \subseteq Q\) a set of qubits which are most likely to have an error based on the log-likelihood ratios. Here \(H_{[R]}\) refers to a restriction of \(H\) to the columns specified in the set \(R\). The OSD decoder then assumes no error on the set of remaining qubits  \(Q \setminus R\) and inverts the truncated matrix to get as a total error guess (up to reordering of the qubits)
\begin{align}
    \vb{x} = (H^{-1}_{[R]} \vb{s}, \vb{0}_{[Q\setminus R]}).
\end{align}
Higher-order OSD improves upon that by considering non-zero configurations \(\vb{x}_{[Q \setminus R]}\) for the remaining qubits and adapting the reliable part to ensure validity as
\begin{align}
    \vb{x} = (H^{-1}_{[R]} \vb{s} + H^{-1}_{[R]} H_{[Q \setminus R]} \vb{x}_{[Q \setminus R]}, \vb{x}_{[Q \setminus R]}).
\end{align}

BP+OSD provides a  a general purpose decoder that is  efficient enough for reasonable benchmarking and has a well tested implementation~\cite{ roffe2020decoding, roffe2020bposd, roffe2022ldpcpython}. In App.~\ref{app:bposd}, we provide details on chosen parameters. 
For small qubit numbers, it is typically observed that the performance of BP+OSD comes close to most-likely error decoding, see App.~\ref{app:bpvalidity} for a comparison. 

\begin{figure}
    \centering
    \includegraphics[width=0.4\textwidth]{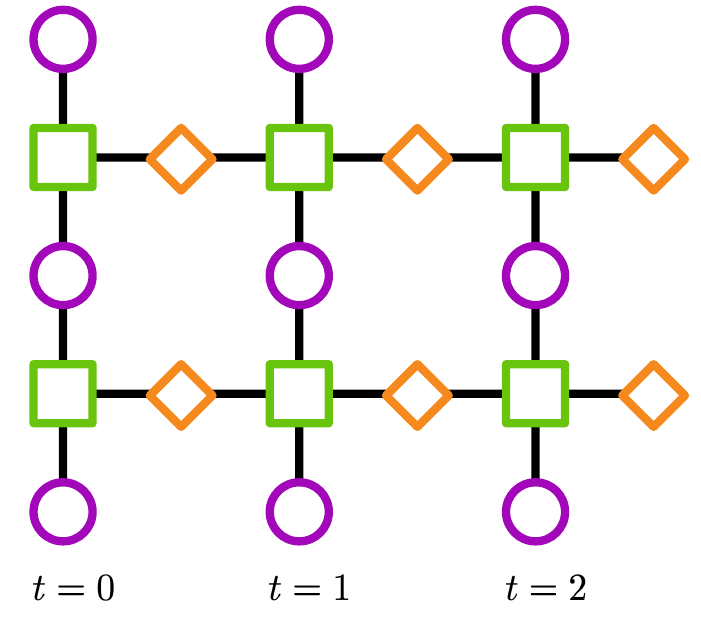}
    \caption{Phenomenological noise Tanner graph for a \(3\)-bit repetition code using \(3\) rounds of noisy measurements. Newly introduced syndrome-error variables \(\tilde{\vb{s}}\) are represented by orange rhombi. Note that this Tanner graph is (up to the rightmost rhombi) equivalent to the Tanner graph of a distance 3 surface code.  }
    \label{fig:rep_code_tg_phenomenological}
\end{figure}

As BP+OSD essentially provides a solution \(\vb{x}\) to \({H \vb{x} = \vb{s} \mod 2}\), we can also use the same algorithm to find a solution to \(A \vb{y} = \Delta \vb{s} \mod 2\) (Eq.~\ref{eqn:phnsto}) representing the phenomenological noise model. In the graphical picture, this corresponds to taking \(d\) copies of the code's Tanner graph, adding new variable nodes for every check and interconnecting them according to the right part of the matrix \(A\). An example of the adapted Tanner graph for a \(3\)-bit repetition code using three rounds of noisy measurements is shown in Fig.~\ref{fig:rep_code_tg_phenomenological}. Note that also here, we simulate a final destructive single qubit readout by a round of noiseless syndrome measurement.

\subsection{Simulation Results}

In the following section, we present results we obtain from sampling LCS codes for code capacity and phenomenological noise channels, which we compare to the performance of the standard surface code. Because LCS and surface codes are both CSS and \(X\)- and \(Z\)-stabilizers are symmetric up to permutations, it suffices to focus on pure bit-flip noise. We start with a discussion of the results for the smallest distance instances of LCS codes. We then move to larger codes, where we first discuss how to compare different finite rate codes and then present the core results regarding logical performance and thresholds of LCS codes.

\subsubsection{Logical error rate and pseudo-threshold}
As we will be considering codes with more than one logical qubit, we will declare logical failure whenever any of the constituent logical qubits has an error. The logical error rate is given by
\begin{align}
    p_{\text{ler}} &= \sum_{w = t+1}^{n} N_{\text{ler}}(w) p^w (1-p)^{n-w} \\
    &= (1-p)^n \sum_{w = t+1}^{n} N_{\text{ler}}(w) \qty(\frac{p}{1-p})^w,
\end{align}
where $N_{\text{ler}}(w)$ is the number of failure configurations of weight $w$ and \(t = \lfloor\frac{d}{2}\rfloor\) is the minimum number of correctable errors. 
Let us start by looking at small LCS codes with \(\ell = 1\), \(L \in \{3,4,5\}\) and parameters \([[15,3,3]], [[20,4,3]]\) and \([[25,5,3]]\) respectively with a rate of \(r = 1/5\). The first two codes are also shown in Fig.~\ref{fig:lts_d3_fam}. The logical error rate against the physical error rate is shown in Fig.~\ref{plot:plot_cc_x_mle}~\((a)\).

We make the following observations about the given distance \(3\) codes: first of all, the different codes show a similar logical error rate across the range of physical error rates, where the logical error rate is larger for larger physical qubit number, as expected when fixing the distance. Furthermore the logical error rate scaling at low $p$ is consistent with \(p_{\text{ler}} \propto p^2\), indicating that arbitrary single qubit errors are corrected. Remarkably, \(3\) copies of distance \(3\) surface codes with parameters \([[39,3,3]]\) have a logical error rate comparable to the \([[20,4,3]]\) LCS code encoding one logical qubit more with fewer physical qubits. 

To further assess the performance of the error correcting codes, we can look at the physically motivated \emph{pseudo-threshold}  \(p^{\star}_{\mathrm{th}}\) \cite{preskill1997fault}. 
It is the value of physical error rate \(p\) below which the logical error rate is lower than the physical error rate. In general, we call the presence of any faulty (logical) qubit in our computational (logical) Hilbert space a (logical) failure. This implies that for $k$ bare physical qubits each failing with \(p\), their total failure probability is given by
\begin{align}
    p_{\text{fail}}(k) &= \sum_{l = 1}^k \binom{k}{l} (1-p)^{k-l} p^l \\
    % &= \sum_{l = 0}^k \binom{k}{l} (1-p)^{k-l} p^l - (1-p)^{k} \\
    % &= (1-p + p)^k  - (1-p)^{k} \\
    &= 1 - (1-p)^{k}. \label{eqn:pfail}
\end{align}
This reduces to the well known case for \(k = 1\), \(p_{\text{fail}} = p\) that is often considered for (planar) color or surface codes hosting one logical qubit.

As shown in Fig.~\ref{plot:plot_cc_x_mle}~\((a)\), The pseudo-thresholds are in the range of \(8-9 \%\).

\subsubsection{Logical error rate per logical qubit}

\begin{figure}
    \centering
    \includegraphics[width=\linewidth]{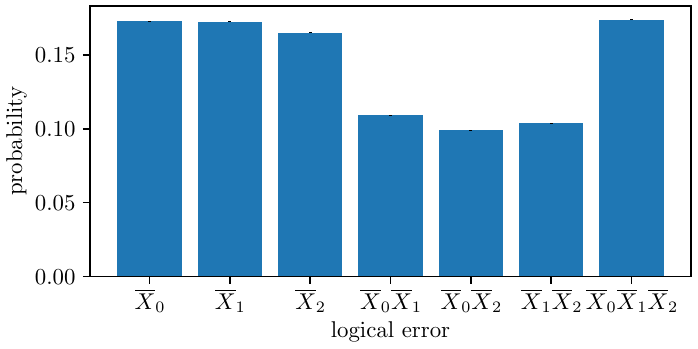}
    \caption{Probability of logical errors at \(p_{\mathrm{x}} = 0.01\) for the \([[15,3,3]]\) LCS code. The logical qubits don't fail independently as \(p(\overline{X}_i) p(\overline{X}_j) \neq p(\overline{X}_i \overline{X}_j)\). Because of the symmetry of the code and the noise, we expect the same logical error rates for e.g.  \(\overline{X}_0\) and \(\overline{X}_2\). We attribute the difference seen above to the specific implementation of the decoder that returns only one solution whenever there are multiple with the same likelihood.}
    \label{plot:logical_correlations}
\end{figure}

In a practical setting, we can not distinguish which of the logical qubits has a logical error and therefore, any error is bad. Additionally, a rescaling of the logical error rate to a \emph{logical error rate per logical qubit} in order to e.g. compare codes with a different number of logical qubits \(k\) is only meaningful if they fail independently of each other. 
Fig.~\ref{plot:logical_correlations} shows the probability of logical errors in the \([[15,3,3]]\) code for a bit-flip error rate $p_{\mathrm{x}} = 0.01$. Most notably, \(p(X_0 X_1 X_2) \neq p(X_0) p(X_1)p(X_2)\) indicates that the logical qubits do not fail independently. This is in contrast to using multiple codes each encoding only a single logical qubit. We will therefore always consider the logical failure of the block as soon as any logical qubit comprising it fails. Whenever we compare to surface codes with \(1\) logical qubit, we obtain the logical error rate for \(k\) copies of surface codes by rescaling the single-logical qubit error rate as
\begin{align}
    p_{\mathrm{L}}^{(k)} = 1 - (1 - p_{\mathrm{L}}^{(1)})^k. \label{eqn:rescaling_ler}
\end{align}

\subsubsection{Copies of single-logical qubit codes or single-block codes?}
In the following, we will discuss and compare logical error rates of various LCS and surface codes encoding the same number of logical qubits \(k=3\) and distance \(d=3\). This corresponds to \(\ell \in \{1,2,3\}\), \(L = 3\). 
Fig.~\ref{plot:plot_cc_x_mle}~\((b)\) shows the logical error rate of these codes sampled over a code capacity channel and decoded with the MLE decoder. For the interconnected \([[39,3,3]]]\)  LCS codes, the logical error rate is almost an order of magnitude lower for small physical bit-flip rates \(p_{\mathrm{x}}\) compared to three copies of \([[13,1,3]]\) surface codes. 
To explain this remember that  at low \(p\) the logical error rate is dominated by low-weight failure configurations. Using a look-up table decoder, we count the number of failure configurations, shown in Tab.~\ref{tab:nfail} for a weight \(w \leq 5\). The number of failure configurations is significantly lower for the LCS code encoding the \(3\) logical qubits in one block which results in the significantly lower logical error rate.
Also the pseudo-threshold is significantly higher with \(\psth([[39,3,3]]-\text{LCS}) \approx 8.6\%\) against \(\psth(3\cdot[[13,1,3]]-\text{S}) \approx 6.4\%\). 
Note that the codes with smaller encoding rate \(r = \frac{k}{n}\) show a lower logical error rate, which is plausible given their correspondingly higher number of stabilizers. However, the \([[75,3,3]]\)-LCS code only has a marginally lower logical error rate compared to the  \([[39,3,3]]\)-LCS code, hinting at a tradeoff between physical qubit number and degeneracy. 
Tab.~\ref{tab:phpsth} summarizes the pseudo-thresholds for these small codes, which are all in the range of \(8 - 9 \%\).

For phenomenological noise, we set the probability of a noisy syndrome measurement to \(q = p_{\mathrm{x}} \ifed p\). Note that slightly different notions of \emph{error correction} and \emph{syndrome measurement cycles} are in use in literature. We follow Refs.~\cite{acharya2023suppressing,bravyi2024high} and report a lower bound on the \emph{logical error rate per syndrome measurement cycle} by rescaling the total logical error rate after \(n_{\mathrm{sc}}\) cycles as 
\begin{align}
    p_{\mathrm{L}}(p) \geq 1 - \left(1 - p_{\mathrm{L}}(n_{\mathrm{sc}})\right)^{\frac{1}{n_{\mathrm{sc}}}}.
\end{align}
This can then be compared to the failure probability of \(k\) unencoded qubits to get the pseudo-threshold as a solution to \(p_{\mathrm{L}}(p) = 1-(1-p)^k\), shown for the MLE decoder and distance \(3\) bounded LCS codes in Fig.~\ref{fig:plot_ph_x_mle}. The pseudo-thresholds are summarized in Tab.~\ref{tab:phpsth} and range between \(4\) and \(5 \%\), higher than the pseudo-threshold of \(3\) copies of distance \(d=3\) surface codes of \(\approx 3 \%\).
Also note that the quadratic scaling in the low error rate regime is consistent with the ability to correct for any order \(p\) error, data qubit or measurement error.

\begin{table}
        \caption{Number of low weight failure configurations of three copies of distance \(3\) surface codes compared to the \([[39,3,3]]\)-LCS code. Up to and including weight \(w = 5\), the interconnected LCS codes have fewer failure configurations, explaining the lower logical error rates.}
        \begin{ruledtabular}
        \begin{tabular}{ccc}
                weight \(w\) & \(3 \cdot [[13,1,3]]\) surface & \([[39,3,3]]\) LCS \\
                \hline
                \(2\) & 69 & 12\\
                \(3\) & 2253 & 852\\
                \(4\) & 34\;\!152 & 23\;\!093\\
                \(5\) & 321\;\!690 & 307\;\!976\\
                \(6\) & 2\;\!163\;\!638 & 2\;\!378\;\!071\\ 
        \end{tabular}
\end{ruledtabular}
\label{tab:nfail} 
\end{table}

\begin{table}
    \caption{Pseudo-thresholds of small LCS codes, compared to three copies of surface codes. Note that the LCS codes have \(26-40 \%\) higher pseudo-thresholds than surface codes for code capacity and \(40-56\%\) higher pseudo-thresholds compared to surface codes with phenomenological noise.}
    \label{tab:phpsth}
    \centering
    \begin{ruledtabular}
    \begin{tabular}{lcc}
        Code & \multicolumn{2}{c}{pseudo-threshold} \\
        & code capacity  & phenomenological \\
        \hline
        \([[15,3,3]]\) LCS     & \(8.1 \pm 0.1 \%\) & \(4.3 \pm 0.1 \%\) \\
        \([[20,4,3]]\) LCS     & \(8.6 \pm 0.1 \%\) & \(4.2 \pm 0.1 \%\) \\
        \([[25,5,3]]\) LCS     & \(8.4 \pm 0.1 \%\) & \(4.2 \pm 0.1 \%\) \\
        \([[39,3,3]]\) LCS     & \(8.7 \pm 0.1 \%\) & \(4.6 \pm 0.1 \%\) \\
        \([[39,3,3]]\) surface & \(6.4 \pm 0.1 \%\) & \(3.0 \pm 0.1 \%\) \\
        \([[75,3,3]]\) LCS     & \(9.0 \pm 0.1 \%\) & \(4.7 \pm 0.1 \%\) \\
    \end{tabular}
    \end{ruledtabular}
\end{table}

\begin{figure*}
    \centering
    \includegraphics[width=\linewidth]{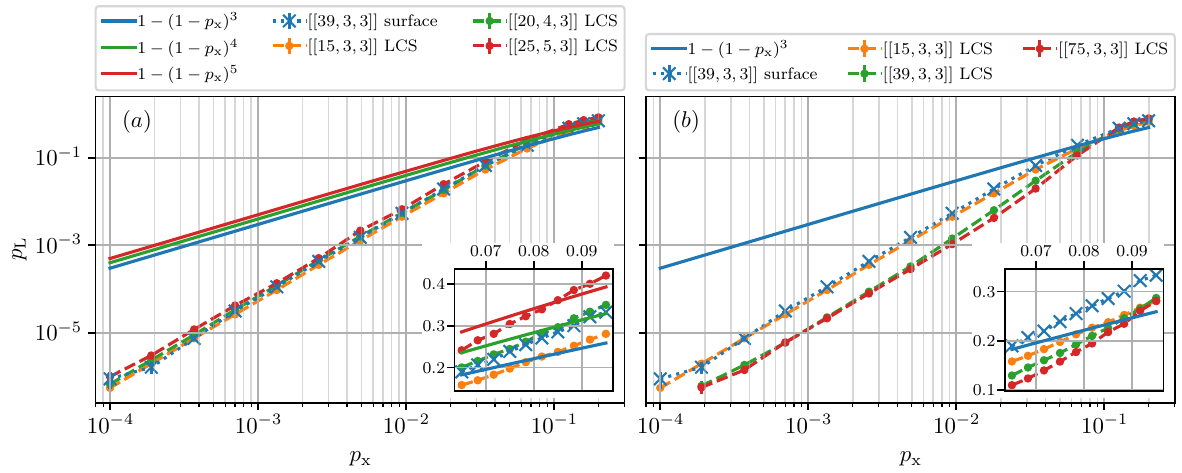}
    \caption{Logical error rates of small LCS and surface codes, sampled over a bit-flip channel and decoded using the MLE decoder (Sec.~\ref{subsec:MLE}).
    \textbf{(a)} Codes with \(\ell = 1\) and \(L = 2,3,4\), compared to error probabilities of \(3,4\) and \(5\) physical qubits (solid lines), respectively.  The smallest distance \(3\) code has the lowest logical error rate. The curves cross the physical qubit curves at the pseudo-threshold \(\psth \approx 8.6\%\).
    \textbf{(b)} Codes with \(\ell = 1,2,3\) and \(L = 3\), compared to error probabilities of \(3\) physical qubits and \(3\) copies of  \([[13,1,3]]\) surface codes.
    Interconnected codes can achieve the same logical error rate as surface codes using less physical qubits (here \(15\) vs. \(39\)). Using more physical qubits increasing the redundancy gives an improvement of almost one order of magnitude for low physical error probabilities. The pseudo-threshold is also increased from  \(\psth(3\cdot[[13,1,3]]-\text{S}) \approx 6\%\) to \(\psth([[39,3,3]]-\text{LCS}) \approx 8.6\%\). In this plot and all following, statistical error bars corresponding to one standard deviation in the Monte Carlo sampling are drawn (occasionally smaller than markers).}
    \label{plot:plot_cc_x_mle}
% \end{figure*}
\hspace{2em}
% \begin{figure*}
    % \centering
    \includegraphics[width=\linewidth]{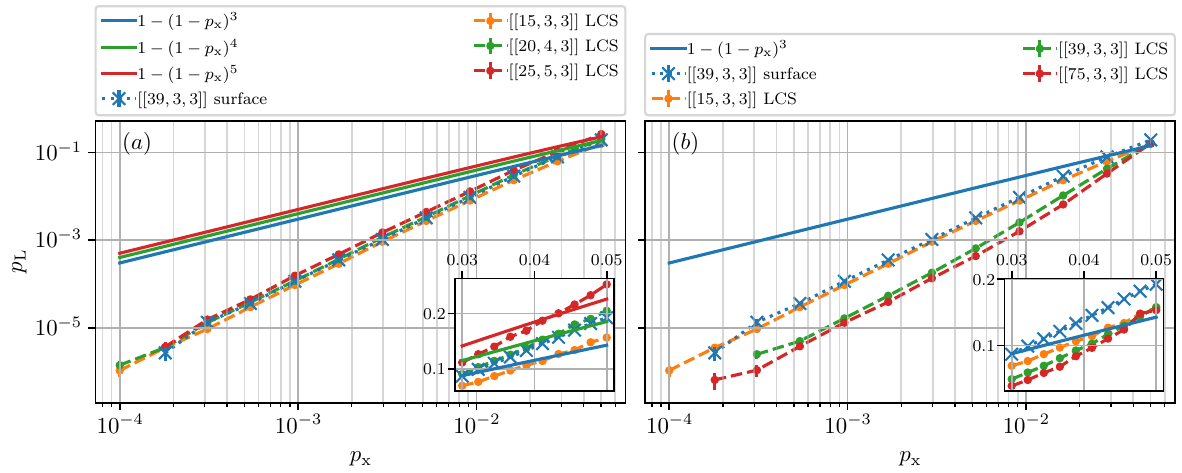}
    \caption{Logical error rates of small LCS and surface codes, sampled over a bit-flip channel with \(d\) noisy syndrome measurements and decoded using the MLE decoder (Sec.~\ref{subsec:MLE}). 
    \textbf{(a)} Codes with \(\ell = 1\) and \(L = 2,3,4\), compared to error probabilities of \(3,4\) and \(5\) physical qubits, respectively.  The smallest distance \(3\) code has the lowest logical error rate. The curves cross the physical qubit curves at the pseudo-threshold \(\psth \approx 4.2\%\).
    \textbf{(b)} Codes with \(\ell = 1,2,3\) and \(L = 3\), compared to error probabilities of \(3\) physical qubits and \(3\) copies of  \([[13,1,3]]\) surface codes.
    Interconnected codes can achieve the same logical error rate as surface codes using less physical qubits (here \(15\) vs. \(39\)). Using more physical qubits increasing the redundancy gives an improvement of almost one order of magnitude for low physical error probabilities. The pseudo-threshold is also increased from  \(\psth(3\cdot[[13,1,3]]-\text{S}) \approx 3\%\) to \(\psth([[39,3,3]]-\text{LCS}) \approx 4.6\%\).}
    \label{fig:plot_ph_x_mle}
\end{figure*}

\subsubsection{Code families and their thresholds} \label{sec:codefamilies}

Beyond pseudo-thresholds of specific codes, the asymptotic \emph{threshold} of families of code is of interest, in particular in order to compare the performance of different code families. 

It is important to note that the existence of a threshold has been proven for concatenated codes~\cite{aharonov2008faulttolerant, knill1998resilient}, codes with an asymptotically constant encoding rate~\cite{kovalev2013fault, gottesman2013fault} and topological codes with rate vanishing \(r \propto n^{-1}\)~\cite{kitaev1997quantum}. For general QLDPC codes, the existence of and lower bounds on a threshold were proven subsequently~\cite{dumer2015thresholds}. 
The threshold of a code family under a specific noise model and decoder is typically determined by Monte Carlo sampling and plotting the logical error rate versus the physical error rate for representatives of the code family with growing number of physical qubits and code distance. Heuristically, a crossing of these curves at a point \(p = p^{\star}\) indicates the existence of a threshold at \(p^{\star}\). 
This is supported by arguments from statistical mechanics mappings of error correcting codes and is particularly successful for topological codes with well defined scaling properties~\cite{Dennis2002, Wang2003, Ohno2004}. Hence for independent copies of surface codes we can expect the same asymptotic threshold as the copies fail independently. Results for \(k=1\) logical qubits that are rescaled according to Eq.~\ref{eqn:rescaling_ler} however have finite size crossing points of the curves that deviate from the asymptotic threshold. For general QLDPC codes, the statistical mechanical picture is more intricate and far less well understood and part of ongoing research~\cite{kovalev2018numerical,rakovszky2023the}.  Nevertheless, crossing points of such curves give an estimate on the value of the threshold. 
In the following, crossing points are obtained from reading off the crossings of quadratic least square fits to the data points. Indicated uncertainties correspond to estimates of how well we can resolve the respective crossing point.

From all available code parameters when sweeping \(\ell,L\) (Fig.~\ref{fig:available_codes_lL}) we can define code families which are suitable for comparative simulations of error correcting properties. In the following, we will define three relevant scenarios and report the numerical findings. 
In the following, we report data for surface codes that is obtained with the same BP+OSD decoder as used for LCS codes. While BP+OSD is a general purpose decoder, surface codes are amenable to tailored decoders, most notably weight matching~\cite{higgott2021pymatching}. Comparing the performance of different decoders can be subtle, however we observed a better logical error rate when decoding with BP+OSD and therefore use the latter for a direct comparison. We discuss this in more detail in App.~\ref{app:bpvalidity}.

\begin{enumerate}
    \item \textbf{Same parameter LCS code}:
    Enforcing the parameters \([[n,k,d]]\) of surface (subindex \(s\)) and LCS codes to be the same corresponds to taking \(L\) copies of distance \(L\) surface codes and interconnecting them. Hence, this scenario explores the influence of the higher stabilizer weights with 2D non-local connectivity. This is achieved by setting \(\ell_s = \ell = L-1 = L_s -1\), such that the parameters are
    \begin{align}
        [[n,k,d]]_{\text{lcs}} = [[n,k,d]]_{\text{s}} = [[(2 L^2 - 2L + 1)L, L, L]].
    \end{align}
     These families are marked dark green in Fig.~\ref{fig:available_codes_lL}.
     
     Under code capacity noise, Fig.~\ref{fig:plot_lcs_cc_bp}~\((a)\) shows the logical error rate of any logical qubit compared to the physical bit-flip probability. With growing \(n,d\), the logical error rate is increasingly suppressed. Compared to copies of surface codes, the interconnected LCS codes show a logical error rate that is up to 2 orders of magnitude lower at physical error rates on the order of \(10^{-2}\).  We observe a  crossing of curves with at \(p^\star \approx 6.5-7\%\) indicating a threshold of \(\approx 6.7 \pm 0.3\%\), slightly lower than the crossing point for surface codes at \(p^\star \approx 7.5 \pm 0.3 \% \). %All indicated uncertainties on threshold values are readoff errors estimating how well we can resolve the respective crossing point. 

    Results using the phenomenological noise model decoded with the adapted BP+OSD decoder are shown in Fig.~\ref{fig:plot_lcs_ph_bp}~\((a)\). The results qualitatively show similar trends as for the previous code capacity case. To pick out a striking example, at a physical error rate of \(10^{-2}\), the logical error rate of the \([[595,7,7]]\) LCS code is 2 orders of magnitude lower compared to the surface code. The estimated threshold value we report for this code family under phenomenological noise is \(\approx 2.9 \pm 0.1\%\) which within resolution coincides with the surface code family threshold \(\approx 2.9 \pm 0.1\%\). 
    
    \item \textbf{Highest rate LCS code} for fixed dimension and distance.  We can see in Fig.~\ref{fig:available_codes_lL} that for the same \(k,d\), we can tune the parameter \(\ell\) for the LCS codes to a higher and lower rate. The highest rate is achieved by setting \(\ell = \frac{L-1}{2}\). These LCS codes are depicted in Fig.~\ref{fig:available_codes_lL} by purple rectangles. This family includes representatives of codes with the smallest number of physical qubits while having distances \(d \geq 3\). 
    We compare this family to surface codes with the same number of logical qubits \(k\) and distance \(d\), which gives 
    \begin{align}
    [[n,k,d]]_{\text{lcs}} &= [[\frac{1}{2}(L^2  + 1)L, L, L]], \\
        [[n,k,d]]_{\text{s}} &= [[(2 L^2 - 2L + 1)L, L, L]].
    \end{align}
    This implies a saving of a factor 4 in qubit number for the same rate and distance for sufficiently large $L$.

    The simulation results under code capacity noise are shown in Fig.~\ref{fig:plot_lcs_cc_bp}~\((b)\). Also in this scenario, we find a lower logical error rate for the LCS family, which can be attributed to the smaller number of physical qubits. However the gain is smaller compared to lower rate codes which can be explained with the higher degeneracy of lower rate codes. We also observe a  crossing of curves at a slightly higher \(p^\star \approx 7.7 \pm 0.2\%\) indicating a threshold in that vicinity.
    
     Under phenomenological noise (Fig.~\ref{fig:plot_lcs_ph_bp}~$(b)$) for the given scenario the crossing indicates a threshold estimate of \(\approx 3.2 \pm 0.1\%\) compared to \(\approx 2.9 \pm 0.1\%\) for the corresponding surface codes. The slight deviation in favor of the LCS family should be interpreted cautiously.
   
    \item \textbf{Largest distance LCS code} for fixed physical and logical qubit number. This compares codes with parameters 
    \begin{align}
    [[n,k,d]]_{\text{lcs}} &= [[\frac{1}{2}(L^2  + 1)L, L, L]], \\
        [[n,k,d]]_{\text{s}} &= [[\frac{1}{2}(L^2  + 1)L, L, \frac{L+1}{2}]].
    \end{align} 
    This implies a factor of 2 improvement in distance for same qubit number and rate for sufficiently large $L$.

    The logical error rates for this scenario under code capacity noise are shown in Fig.~\ref{fig:plot_lcs_cc_bp}~\((c)\) and under phenomenological noise in Fig.~\ref{fig:plot_lcs_ph_bp}~\((c)\). Here, we observe the largest gain among the three scenarios, i.e. when fixing the number of physical qubits and the desired code rate, choosing LCS codes over surface codes leads to the largest improvement in logical error rate. This can likely be attributed to the fact that the corresponding surface codes have a smaller distance \(d_{\mathrm{s}} = \frac{d_{\mathrm{lcs}}+1}{2}\). Compared to the example in the "same parameter scenario", here we observe an improvement of 2 orders of magnitude at physical error rate $p=10^{-2}$ already for the even smaller \([[369,9,9]]\) LCS code compared to the corresponding nine copies of distance five surface codes with parameters \([[369,9,5]]\). Note that the threshold values for this scenario are the same as in the previous scenario.
    
    \end{enumerate}

Overall, these results indicate that LCS codes systematically offer improved performance in terms of logical error rate compared to standard surface codes under both code capacity and phenomenological noise.

\begin{figure*}
    \centering
    \includegraphics[width=\linewidth]{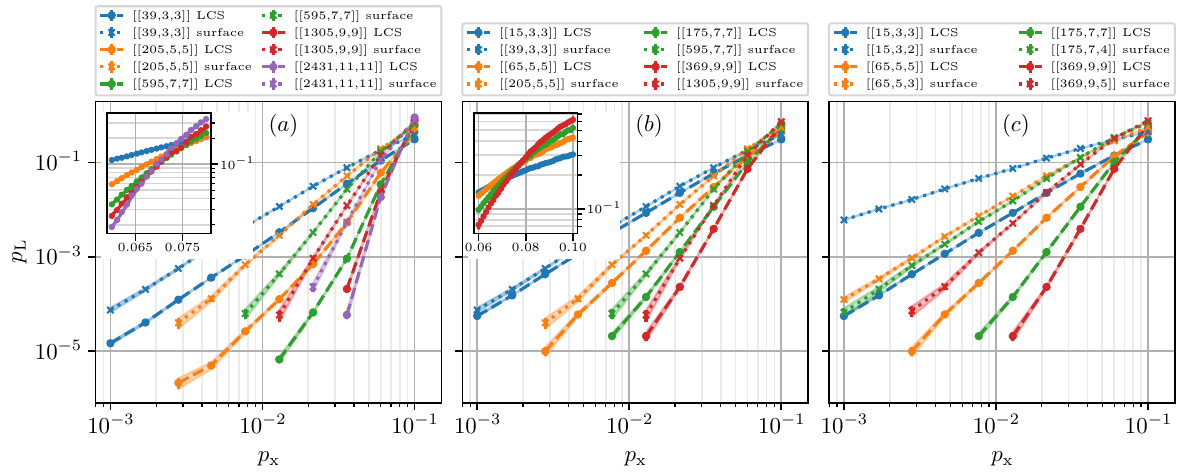}
    \caption{Logical error rate of LCS codes under \emph{code-capacity bit-flip noise} using the BP+OSD decoder compared to surface codes. We compare three different parameter scaling choices, for all of which we observe a reduction of logical error rate with growing \(n,d\).  
    \textbf{(a)} Codes are chosen such that LCS and surface codes have the same parameters \([[n,k,d]]\). The logical error rate for the interconnected codes is up to 2 orders of magnitude lower than for the surface codes at a physical error rate in the order of \(10^{-2}\). We also observe a crossing of the curves indicating the existence of a threshold at \(\approx 6.7 \pm 0.3 \%\) . 
    \textbf{(b)} LCS codes are chosen such that they have the lowest qubit number \(n\) for same \(k,d\). As explained in the text, there is still a reduction in logical error rate compared to disjoint surface codes, however this is less pronounced than in the setting discussed previously. The crossing of the curves indicates a slightly higher threshold at \(\approx 7.7 \pm 0.2\%\). 
    \textbf{(c)} Codes chosen such that for the same \(n,k\), LCS codes have highest available distance. Here we observe the largest reduction in logical error rate. Note that this is consistent with the expectation that the distance is closely related to the number of correctable errors.}
    \label{fig:plot_lcs_cc_bp}
% \end{figure*
\hspace{0.5em}
% \begin{figure*}
    % \centering
    \includegraphics[width=\linewidth]{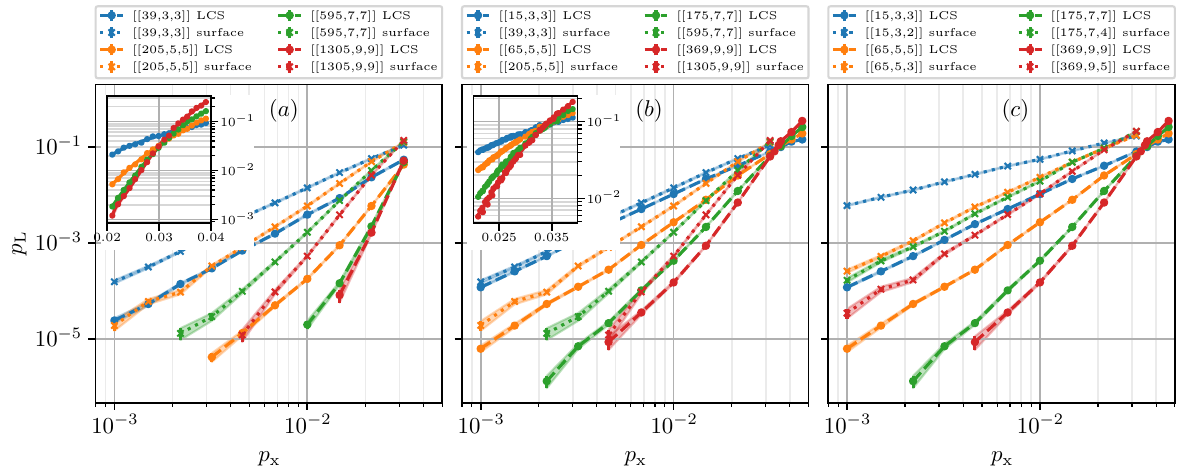}
    \caption{Logical error rate of LCS codes under \emph{phenomenological bit-flip noise} using the BP+OSD decoder compared to surface codes. We simulate $d$ cycles of syndrome measurements and report the logical error rate per cycle. We compare three different parameter scaling choices, for all of which we observe a reduction of logical error rate with growing \(n,d\).  
    \textbf{(a)} Codes are chosen such that LCS and surface codes have the same parameters \([[n,k,d]]\). The logical error rate for the interconnected codes is up to 2 orders of magnitude lower than for the surface codes at a physical error rate in the order of \(10^{-2}\). We also observe a crossing of the curves in the vicinity of \(\approx 2.9 \pm 0.1 \%\) indicating the existence of a threshold. 
    \textbf{(b)} LCS codes are chosen such that they have the lowest qubit number for same \(k,d\). As explained in the text, there is still a reduction in logical error rate compared to disjoint surface codes, however not as large. The crossing of the curves indicates a slightly higher threshold at  \(\approx 3.2 \pm 0.1\%\). 
    \textbf{(c)} Codes chosen such that for the same \(n,k\), LCS codes have highest available distance. Again, note that this is consistent with the expectation that the distance is closely related to the number of correctable errors.}
    \label{fig:plot_lcs_ph_bp}
\end{figure*}

\section{Fault tolerant syndrome measurement circuits} \label{sec:ft}
The simulations with a phenomenological noise model show the high intrinsic error correction capabilities of LCS codes compared to surface codes, even under erroneous syndrome measurements. It is, however, not clear how this translates to an implementation with full circuit level noise because of the weight-6 instead of weight-4 stabilizers. Coupling in ancillary qubits increases the number of failure configurations and can even reduce the effective distance of the code. A noteworthy example are hook errors in topological color codes. Circuits have to be designed carefully, such that faults occurring during the syndrome measurement do not spread in an uncontrollable way~\cite{landahl2011fault}. 

For HGP codes, it was shown that any order of two qubit gates does not spread errors catastrophically, because the relevant stabilizers overlap with any logical operator on one location only~\cite{manes2023distance}, which is a sufficient condition. For (non-rotated) surface codes constructed as HGP of two classical repetition codes, the order does therefore also not matter.
The argument of Ref.~\cite{manes2023distance} does not generalize to LP codes. This can be seen in Fig.~\ref{fig:pcm_blockstructure}, where the first $Z$-stabilizer overlaps with the first logical $Z$-operator on two locations.

Remarkably, we find that syndrome measurement circuits based on the \emph{coloration circuit} of Ref.~\cite{tremblay2022constant} are distance preserving. We generate the required edge coloring for parallelizing entangling gates according to Alg.~\ref{alg:coloring}. 
Since the maximum degree of the edges in the Tanner graphs of LCS codes is \(\Delta = 6\), this algorithm colors the edges of the Tanner graphs of LCS codes in \(6\) colors~\footnote{We implement the edge coloring using the \texttt{maximal\_matching} algorithm of the \texttt{NetworkX} \texttt{python} library.\cite{hagberg2008exploring}}. 
From this coloring, we construct circuits as shown in Alg.~\ref{alg:circuit} and implement them using \texttt{stim}~\cite{gidney2021stim}. One syndrome measurement cycle of the \([[15,3,3]]\) code is shown in Fig.~\ref{fig:circuit_15_3_3}.
We implement a circuit level noise model where each circuit element fails with probability \(p\), i.e. after single qubit operations, a single qubit depolarizing error \(E \in \{X,Y,Z\}\) occurs with probability $p$. After two qubit entangling operations, one of fifteen  \(E \in \{I,X,Y,Z\}^{\otimes 2} \setminus \{I,I\}\) is placed with probability \(p\).
Initialization and measurements (of ancilla and data qubits) are also noisy, modeled by single qubit bit (phase) flips with probability $p$ after the $Z$ ($X$)- initialization or before the $Z$ ($X$)-measurement. Idling positions suffer from uniform depolarizing noise with $p_{\mathrm{idle}} = p/10$~\footnote{We use $p/10$ instead of $p$ to account for the optimization potential in parallelizing the circuits.}.

The resulting circuits preserve the distance of the static code in the sense that it requires $d$ circuit level faults (including two qubit errors after entangling gates) to cause a logical error without violating a stabilizer~\footnote{This can be done in \texttt{stim} using \texttt{circuit.search\_for\_undetectable\_logical\_errors($\dots$)}}.

\begin{figure*}
    \centering
    \includegraphics[width=\linewidth]{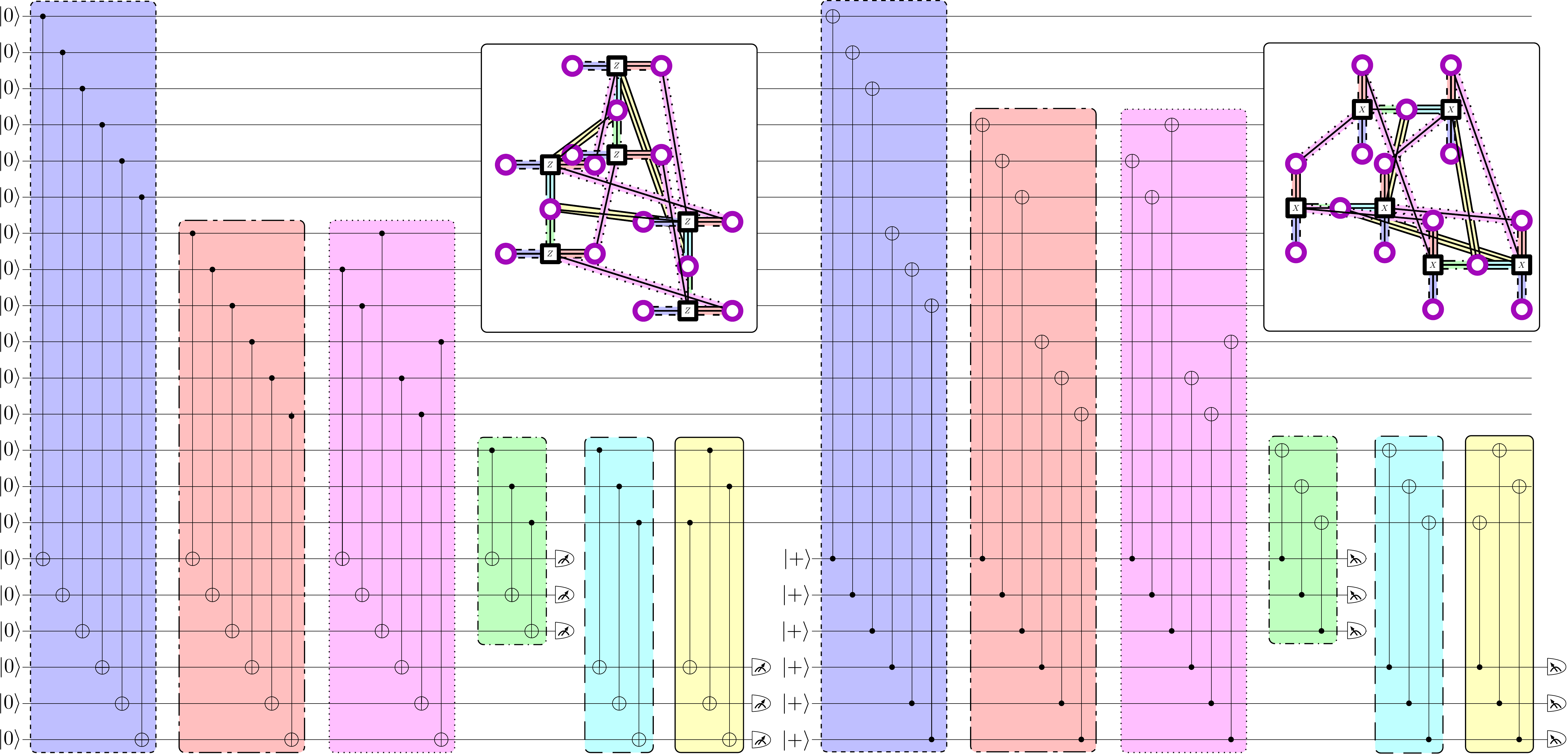}
    \caption{A syndrome measurement cycle for the $[[15,3,3]]$-LCS code constructed using the coloration circuit as described in the main text. $Z$- and $X$-stabilizer measurements are performed one after another in $6$ layers respectively, showing potential for further parallelization. Inset are the $Z$- and $X$-Tanner graphs with edges backed with the same colored and shapes as the corresponding parallel blocks of $\CNOT$s in the circuit. }
    \label{fig:circuit_15_3_3}
\end{figure*}

We can then benchmark the memory capabilities of the LCS codes by sandwiching the coloration syndrome readout circuits with initialization in $\ket{0}^{\otimes n} (\ket{+}^{\otimes n})$ and single-qubit measurement $M_Z^{\otimes n} (M_X^{\otimes n})$ for the $Z(X)$-logical operators. Both these operations are also assumed noisy (with probability $p$), modeled by single qubit bit (phase) flips after the $Z(X)$- initialization or before the $Z(X)$-measurement .

For decoding, we generate a circuit level parity check matrix. The rows represent \emph{detectors}, i.e. sets of deterministic measurements such as two consecutive stabilizer measurement outcomes. The columns represent \emph{error mechanisms} ~\footnote{To generate the decoding matrix, we use \texttt{stim}'s automated detector error model generation.}. We finally decode using BP+OSD similar to the code capacity and phenomenological noise model.

Fig.~\ref{fig:plot_circuit_level} (a) shows the logical error rate per syndrome measurement cycle after $d$ cycles. We show \([[15,3,3]]\),  \([[39,3,3]]\) and  \([[65,5,5]]\) LCS codes compared to $d$ copies of distance \(3, 5\) and \(7\) surface codes. 
This data verifies the fault-tolerance of the circuits by showing a scaling of \(p_L \propto p^{t+1}\) in the regime of low physical error rates.
Sub-threshold and within error bars, the logical error rates of the \([[15,3,3]]\) and \([[65,5,5]]\) LCS codes are the same as their surface code counterparts with parameters \([[39,3,3]]\) and \([[205,5,5]]\). LCS codes can therefore reach the same logical error rate using \(2.78\) and \(3.24\) times fewer physical qubits. 
Again, encoding \(3\) logical qubits using more redundancy, i.e. $39$ physical qubits decreases the logical error rate by almost an order of magnitude.

Fig.~\ref{fig:plot_circuit_level} (b) shows that pseudo-thresholds of LCS codes are significantly higher, \(\psth([[15,3,3]]) \approx 0.45\%\) and  \(\psth(([[39,3,3]]) \approx 0.5\%\), compared to surface codes with \(\psth([[39,3,3]]) \approx 0.39\%\). This is a strong indication that already at small qubit numbers, even under circuit level noise, LCS codes can outperform surface codes.

Even if phenomenological noise thresholds are similar, the (asymptotic) value of the circuit level threshold is typically observed to be reduced for higher stabilizer weights~\cite{ruiz2024ldpc}.
We observe this in Fig.~\ref{fig:plot_circuit_level} (c), showing the crossing of logical error rates for members of the highest rate LCS codes. Their crossing point is at $\pth \approx 0.5\%$ compared to $\pth \approx 0.9\%$ for surface codes, indicating a threshold in that vicinity. 

\begin{algorithm}
        \SetAlgoLined
        \LinesNumbered
        \DontPrintSemicolon
        \BlankLine
        \KwIn{$X$- or $Z$- Tanner graph $T$ of LCS code $\mc{Q}$}
        \KwOut{Edge coloring $C$}
        $C = []$ (empty list) \;
        \While{$T$ is not empty}{
        $m = $ maximal matching of edges in $G$ \;
        $C$ $\leftarrow$ $[m]$ \;
        remove edges $m$ from $T$ 
        }
        \KwRet{\(C\)}
        \caption{Edge coloring of Tanner graph of LCS codes}
        \label{alg:coloring}
\end{algorithm}

\begin{algorithm}
        \SetAlgoLined
        \LinesNumbered
        \DontPrintSemicolon
        \BlankLine
        \KwIn{$X$- and $Z$- edge colorings $C_X, C_Z$ of Tanner graphs of $(\ell,L)$-LCS code $\mc{Q}$}
        \KwOut{Syndrome readout circuit}
        \For{Cycle in range($L$)}{
         Initialize $\frac{n-k}{2}$ $Z$-ancilla qubits in $\ket{0}$\;
         \For{color $c$ in $C_Z$}{
          \For{edge $e = (d,a)$ in color $c$}{
           $\CNOT_{d \to a}$
          }
         }
        Measure $\frac{n-k}{2}$ $Z$-ancilla qubits in $Z$-basis\;
        Initialize $\frac{n-k}{2}$ $X$-ancilla qubits in $\ket{+}$\;
         \For{color $c$ in $C_X$}{
          \For{edge $e = (d,a)$ in color $c$}{
           $\CNOT_{a \to d}$
          }
         }
        Measure $\frac{n-k}{2}$ $X$-ancilla qubits in $X$-basis\;
        }
        
        \KwRet{\(C\)}
        \caption{Coloration syndrome readout circuits for LCS codes}
        \label{alg:circuit}
\end{algorithm}

\begin{figure*}
    \centering
    \includegraphics[width=\linewidth]{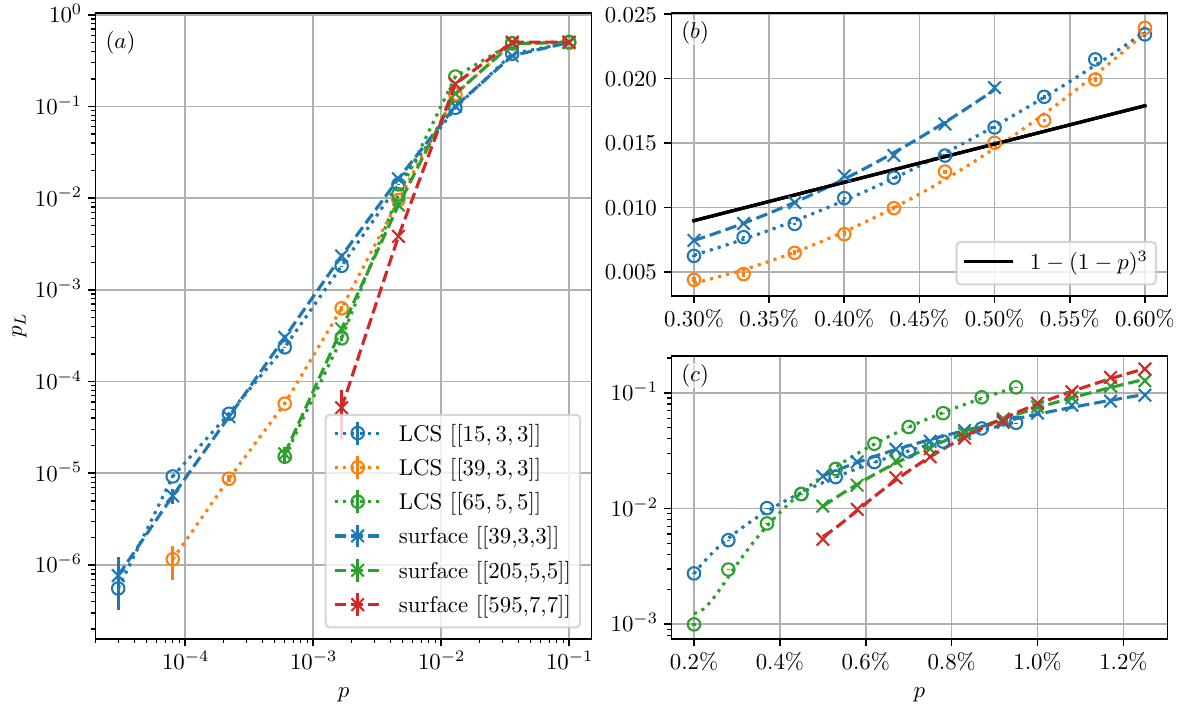}
    \caption{Logical error rates of LCS codes under \emph{circuit level noise}, decoded using BP+OSD, compered surface codes decoded using a minimum weight perfect matching decoder. We simulate $d$ syndrome measurement cycles and report the logical error rate per cycle. \textbf{(a)} Distance $3$ and $5$ LCS codes show the expected scaling $p_L \propto p^2$ and $p_L \propto p^3$ for small $p$, confirming the fault-tolerance property of the syndrome readout circuit implementations. Sub-threshold and within error bars, the logical error rates of the \([[15,3,3]]\) and \([[65,5,5]]\) LCS codes are the same as their surface code counterparts with parameters \([[39,3,3]]\) and \([[205,5,5]]\). \textbf{(b)} Pseudo-thresholds of LCS codes are significantly higher, \(\psth([[15,3,3]]) \approx 0.45\%\) and  \(\psth(([[39,3,3]]) \approx 0.5\%\), compared to surface codes with \(\psth([[39,3,3]]) \approx 0.39\%\). \textbf{(c)}  The crossing point of highest rate LCS codes is at $\pth \approx 0.5\%$ compared to $\pth \approx 0.9\%$ for surface codes, indicating threshold in that vicinity.}
    \label{fig:plot_circuit_level}
\end{figure*}

\section{Towards logical gates in LCS codes} \label{sec:towards}
Implementing logical gates in QLDPC codes is a hard challenge, in particular if the implementation is to be fault-tolerant. In the following, we show a particular set of fault-tolerant gates for LCS codes when fixing \(\ell = 1\). These have an almost planar geometrical representation and useful symmetries. Three representatives (\(L = 2,3,4\)) are shown in Fig.~\ref{fig:lts_d3_fam}. They correspond to codes with parameters \([[n,k,d]] = [[10,1,2]], [[15,3,3]]\) and \([[20,4,3]]\) respectively. They have a symmetry also called \(ZX\)-duality that exchanges \(X\)- and \(Z\)- stabilizers by simple swaps but keeps logical operators invariant. 
This symmetry is easily seen in the parity check matrices for \(\ell = 1\),
\begin{align}
    \tilde{H}_X &= \mqty(1 & 1+P^{(1)} & 0 & 0 & 1 \\
                        0 & 0 & 1 & 1+P^{(1)} & 1+P^{(1) T}) \\
    \tilde{H}_Z &= \mqty(1 & 0 & 1+P^{(1)} & 0 & 1 \\
                        0 & 1 & 0 & 1+P^{(1)} & 1+P^{(1) T}).
\end{align} 
Swapping the second and third block, i.e.~applying the column permutation
\begin{align}
    \tau = \mqty(L&2L)\mqty(L+1&2L +1) \dots \mqty(2L-1&3L-1)
\end{align}
brings \(\tilde{H}_X \to \tilde{H}_Z\) and vice versa. Logical operators of these codes are given by
\begin{align}
    \tilde{L}_X &= \mqty(1 & 0 & 0 & P^{(1)} & 1) \\
    \tilde{L}_Z &= \mqty(1 & 0 & 0 & P^{(1)} & 1)
\end{align} 
and therefore left invariant under \(\tau\).

This duality enables several \emph{fold-transversal} gates, in particular a fold-transversal Hadamard that implements a global logical Hadamard~\cite{breuckmann2022fold},

\begin{align}
    \overline{H} = \bigotimes_{i < \tau(i)} \SWAP(i,\tau(i)) \bigotimes_{i=1}^n H_i = \bigotimes_{i=1}^k\overline{H}_i
\end{align}

and a fold-transversal Phase gate 
\begin{align}
        \overline{S} = \bigotimes_{i \in A} S_i \bigotimes_{i \in B} S^\dagger_i \bigotimes_{\substack{i = 1 , \dots , n \\ i < \tau(i)}} \mathrm{CZ}_{i, \tau(i)}. \label{eqn:phase_gate}
\end{align}
In the latter, \(A\) and \(B\) are partitions of the qubits not involved in \(\tau\) such that every \(X\)-stabilizer has half support on \(A\) and \(B\) respectively. These are also indicated by empty and filled circles in Fig.~\ref{fig:lts_d3_fam} for the \([[15,3,3]]\)-LCS code. 
The \(\mathrm{CZ}\)s ensure that the stabilizer group is invariant by mapping \(S_X \to S_X S_Z\) since for single-qubit Paulis \(\mathrm{CZ}_{ij} X_i \mathrm{CZ}_{ij} = X_i Z_j\). The phase gate acts as \(S_i X_i S_i^\dagger = i X_i Z_i\) and \(S_i^\dagger X_i S_i = -i X_i Z_i\), such that the partitioning guarantees unwanted phase factors to cancel out. We can verify that the fold-transversal Phase gate implements a global logical Phase gate \(\overline{S} = \otimes_{i=1}^k\overline{S}_i\).

These gates can be supplemented by the transversal \(\overline{\mathrm{CNOT}}\)~\cite{shor1996fault} to generate the full Clifford group on the set of logical qubits of copies of these \(d=3\) LCS codes. A universal set of fault-tolerant gates operating on single logical qubits is an open challenge. Approaches towards that goal may be based on (generalized) lattice surgery and gate teleportation \cite{horsman2012surface,brun2015teleportation, cohen2022low}, pieceable fault-tolerance~\cite{yoder2016universal} or code switching~\cite{beverland2021cost,butt2023fault}. Generalizing existing approaches for HGP codes~\cite{krishna2021fault,quintavalle2021single} to LP codes are another promising avenue. 

\begin{figure}
    \centering
    \includegraphics[width=0.48\textwidth]{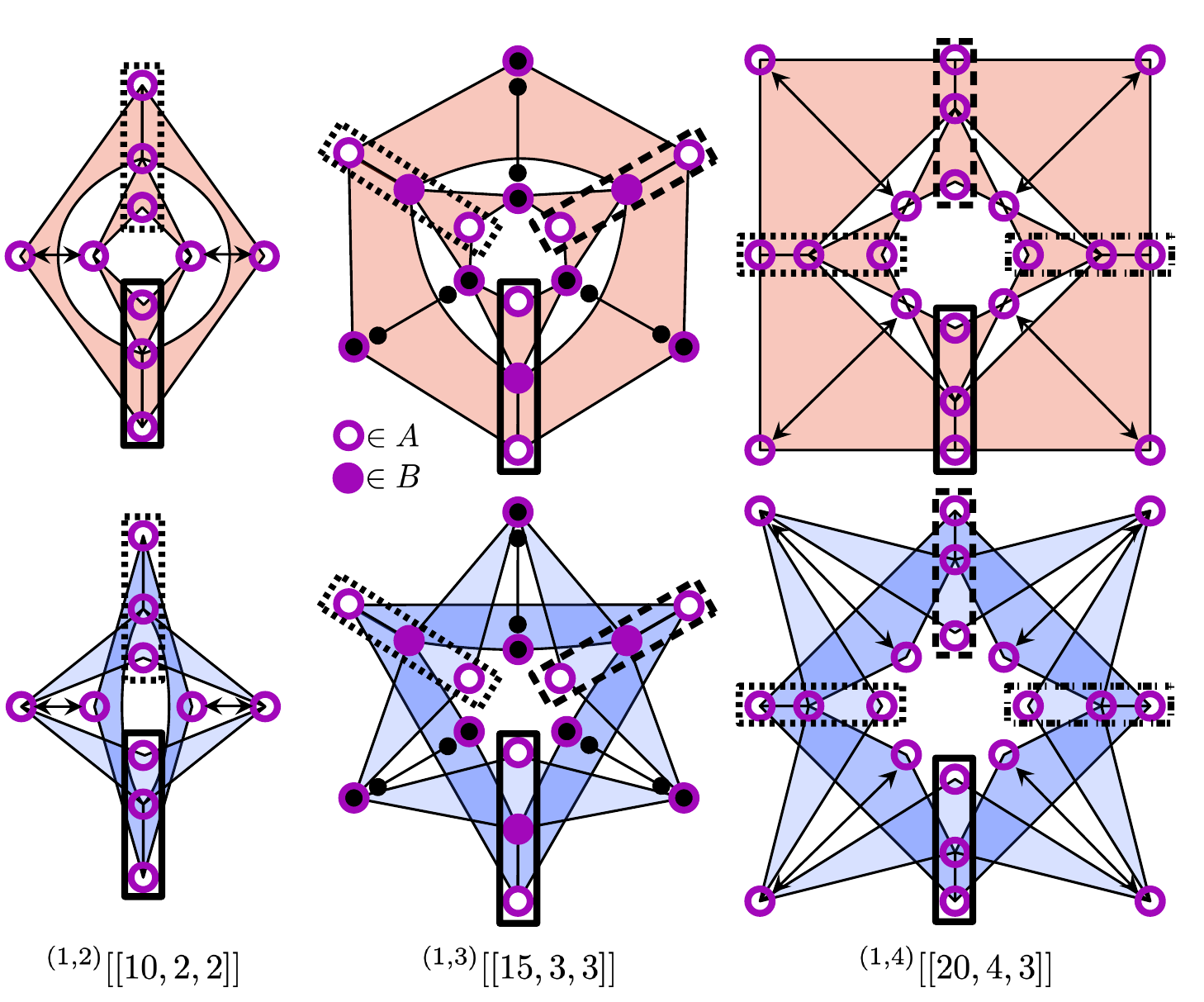}
    \caption{First three members of \(\ell = 1\) LCS codes. \(X\)- (top) and \(Z\)-stabilizers (bottom). Logical operators shown in squares with different line styles for \(X\) and \(Z\) logical operators respectively. Note that they can be chosen to have same support. Swapping pairs of qubits on a diagonal as shown exemplarily for the \(k=2\) and \(k=4\) codes swaps \(X\) and \(Z\) stabilizers while leaving logical operators invariant. For \(k=3\) (center), the partitions \(A\) and \(B\), as well the the \(\mathrm{CZ}\) operations of the fold-transversal Phase gate are indicated (cf. Eq.~\ref{eqn:phase_gate}).}
    \label{fig:lts_d3_fam}
\end{figure}

\section{Conclusion} \label{sec:conclusion}
We have introduced lift-connected surface (LCS) codes as a new family of QLDPC codes based on the lifted product construction. We have shown how they can be viewed as interconnected copies of surface codes and how this additional connectivity leads to favorable parameters compared to disjoint copies of surface codes. . 
We adapted a BP+OSD decoder to the phenomenological noise setting and performed benchmarking of LCS codes for code-capacity and phenomenological noise. We observed that, while asymptotic thresholds are comparable to those of standard surface codes, the pseudo-thresholds can be significantly higher and logical error rates much lower. 
These advantages in particular also hold for small codes with qubit number below 100.

Implementing the stabilizer measurements of the codes in general involves coupling the data qubits to ancilla qubits by gates.
These introduce another source of noise, that is captured by a \emph{circuit level} noise model. 
Investigating the performance of LCS codes under circuit level noise will hinge on the construction of fault-tolerant circuits. For hypergraph product codes, constructions of distance-preserving circuits have been recently introduced~\cite{tremblay2022constant,xu2023constant,manes2023distance}, a generalization to lifted product codes would be highly desirable. 
We have constructed fault-tolerant circuits for three members of the LCS code family that already demonstrate properties outperforming copies of surface codes. They also show potential for further improvement, for example by interleaving $X$- and $Z$- stabilizer measurements.
Other modern fault-tolerant circuit constructions such as using flag-qubits~\cite{chamberland2018flag} should facilitate resource efficient and fault-tolerant stabilizer readout protocols. The circuit construction also depends on assumptions on available gates and connectivity. 
Given that LCS codes are embeddable in three dimensions with local connectivity, this makes them highly attractive for near-term platforms and particularly suited for the emerging platforms of static 3D optical lattices or reconfigurable 2D arrays of Rydberg atoms~\cite{barredo2018synthetic,Bluvstein2023}.

\section*{Acknowledgements}
This research is part of the Munich Quantum Valley (K-8), which is supported by the Bavarian state government with funds from the Hightech Agenda Bayern Plus.
We additionally acknowledge support by the BMBF project MUNIQC-ATOMS (Grant No. 13N16070).
The authors gratefully acknowledge funding by the Deutsche Forschungsgemeinschaft (DFG, German Research Foundation) under Germany’s Excellence Strategy ‘Cluster of Excellence Matter and Light for Quantum Computing (ML4Q) EXC 2004/1’ 390534769. 
Furthermore, we receive funding from the European Union’s Horizon Europe research and innovation programme under grant agreement No. 101114305 (“MILLENION-SGA1” EU Project) and ERC Starting Grant QNets through Grant No. 804247.
M.M. also acknowledges support for the research that was sponsored by IARPA and the Army Research Office, under the Entangled Logical Qubits program through Cooperative Agreement Number W911NF-23-2-0216.
The views and conclusions contained in this document are those of the authors and should not be interpreted as representing the official policies, either expressed or implied, of IARPA, the Army Research Office, or the U.S. Government. 
The U.S. Government is authorized to reproduce and distribute reprints for Government purposes notwithstanding any copyright notation herein.
The authors gratefully acknowledge the computing time provided to them at the NHR Center NHR4CES at RWTH Aachen University (Project No. p0020074). This is funded by the Federal Ministry of Education and Research and the state governments participating on the basis of the resolutions of the GWK for national high performance computing at universities.

\clearpage

\appendix
\section{Block structure of parity check matrices and distance of LCS codes} \label{app:pcm}
The block structure of the parity check matrices of the LCS codes is shown in Fig.~\ref{fig:pcm_blockstructure}, exemplarily for \((\ell,L) = (2,4)\) (\([[52,4,4]]\)). The parity check matrices inherit the block structure from the HGP, which can for example be seen in the block-diagonal structure of \(H_X\). In total, the binary parity check matrices consist of  \(\ell(\ell + 1 )\times (2 \ell^2 + 2 \ell + 1)  \) blocks of size \(L \times L\). Note that both \(H_X\) and \(H_Z\) are in row echelon form.

\begin{figure*}
        \centering
        \includegraphics[width=0.9\linewidth]{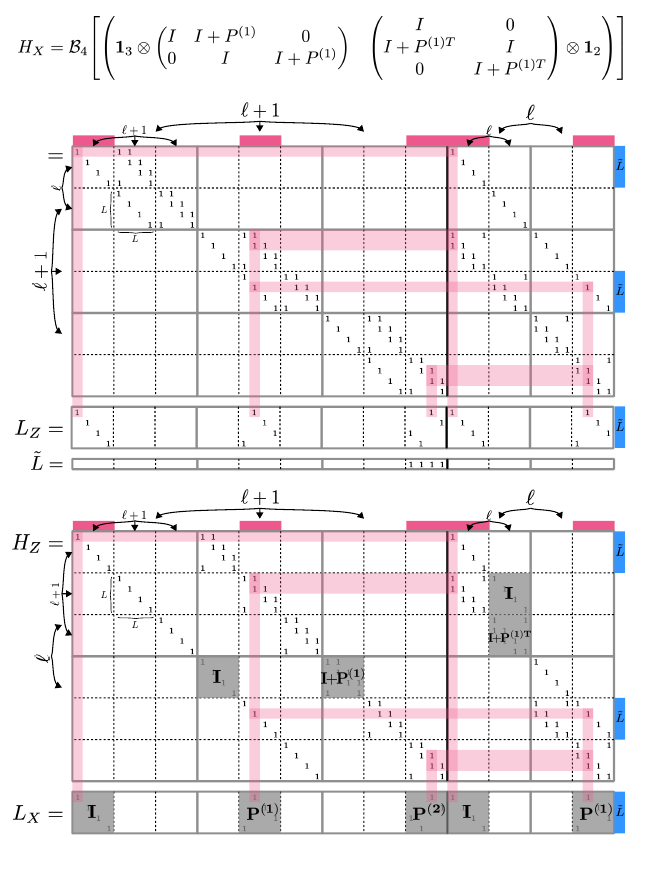}
        \caption{Block structure of the parity check matrices of \(X\)- (Top) and \(Z\)-stabilizers (bottom), exemplarily for \((\ell,L) = (2,4)\). Note that the block structure is inherited from the HGP product. The red rectangles on top of the matrices indicate the positions of the diagonal logicals in the regular HGP surface code.
        Also shown are matrices \(L_Z\) and \(L_X\) that are the binary representation of logical \(Z\)- and \(X\)-operators. They can be found as described in the text and have weight \(2 \ell + 1\). For the first logical operators, the overlap with the stabilizers of the respective other Pauli type is drawn in a red shade, visualizing the commutation. Additionally, we show the operator \(\tilde{L}\) of weight \(L\) which can be constructed from products of all logical operators and some stabilizers as described in the text and shown with blue rectangles on the right side of the matrices.}
        \label{fig:pcm_blockstructure}
\end{figure*}

This particular form of the parity check matrices allows us to construct set of logical operators. If we have a set of \(k\) independent generators of logical operators and find their minimal weight, we can find the minimum distance of the code. 
By carefully inspecting the parity check matrices, we explicitly construct \(L=k\) logical operators of weight \(2\ell + 1\).
To that end, it is useful to first realize that, in regular surface codes, we can choose representatives of logical \(X\)- and \(Z\)- operators to have the same support by putting them on the diagonal. 
Indexing each column of the left with tuples \((i,j)\in \{0,1,\dots,\ell\}^{2}\) and for the right block with   tuples \((i,j) \in \{0,1,\dots,\ell-1\}^{2}\), these diagonal qubits can be identified with columns \((0,0),(1,1),(2,2),\dots,(\ell,\ell)\) in the left part of the parity check matrices and columns \((0,0),(1,1),(2,2),\dots,(\ell-1,\ell-1)\) of the right part. These positions are also indicated by red boxes at the top of the parity check matrices in Fig.~\ref{fig:pcm_blockstructure}. The lift requires to select circulants at these positions, such that the resulting operators are also logical operators. It turns out that choosing logicals (before lifting \(L\) times) of the form
\begin{widetext}
\begin{align}
    \big(\underbrace{\underbrace{\mqty{1 & 0 & \cdots & 0}}_{\ell + 1} \vert \mqty{ 0 & P^{(1)} & 0 & \cdots & 0} \vert  \mqty{0 & 0 & P^{(2)} & 0 & \cdots & 0} \vert \cdots \vert \mqty{0 & \cdots & 0  & P^{(\ell)}}}_{\ell + 1} \vert \vert  \underbrace{\underbrace{\mqty{1 & 0 & \cdots & 0}}_{\ell} \vert \mqty{ 0 & P^{(1)} & 0 & \cdots & 0} \vert \cdots \vert \mqty{0 & \cdots & 0  & P^{(\ell-1})}}_{\ell}   \big)
\end{align}
\end{widetext}
ensures that every stabilizer has even overlap with the logicals. Note that in column \((i,i)\) (of both left and right part), the circulant \(P^{(i)}\) is placed.
These are also shown in Fig.~\ref{fig:pcm_blockstructure} for the \([[52,4,4]]\) code.
As binary representations of logical \(X\)- and \(Z\)-operators, these are \(L\) pairs of disjoint operators since they only consist of circulants with one term, i.e. cyclic permutation matrices. With the respective partner (\(L_X^i, L_Z^i\)), the anti-commutation is guaranteed by the odd weight \(2 \ell + 1\).
Finally, these operators cannot have their weight reduced by adding stabilizers, which can also be verified using the block-structure of the parity check matrices. Every attempt to reduce the weight necessarily introduces new qubit connectivity.

However, taking the product of all these \(L\) operators results in an operator with ones in all the columns specified above. Multiplying stabilizers of rows \((0,0),(1,1),\dots,(\ell-1,\ell-1)\) will give an operator of (potentially lower) weight \(L\). 

We therefore found a set of \(L\) independent operators, each of minimum weight \(2 \ell + 1\). Their sum has minimum weight \(L\) and all other combinations of stabilizer and logical operator have weight \(\geq  2 \ell + 1\) . We have therefore further evidence that the minimum distance of the code is
\begin{align}
    d = \min(2 \ell + 1,L).
\end{align}

\section{Parameters for BP+OSD decoding} \label{app:bposd}
The BP+OSD decoder used in Sec.~\ref{sec:Noisy} has a range of parameters that influence the decoding performance. For a comprehensive overview, refer to the source code~\cite{roffe2020bposd}. Important parameters used in these simulations are shown in Tab.~\ref{tab:bp_params}. We observed that the (standard, but more complex) \emph{product sum} method of BP (also described in the text) performs better than the lower complexity \emph{minimum sum} method also provided by the package. The maximum number of BP iterations and the OSD order are also set heuristically based on observations in the decoding. For the \([[369,9,9]]\)-LCS code, logical error rates for code capacity noise and different numbers of BP iterations are shown in Fig.~\ref{plot:bposd_9_maxiter}. For different codes, we show the logical error rate for different numbers of BP iterations at a fixed physical error rate \(p = 0.0129\) in Fig.~\ref{plot:bposd_maxiter}. Heuristically, we find good performance if we set the maximum number of BP iterations to \(\lfloor\frac{d}{2}\rfloor\).
We attribute this to the observation that if BP is not successful, a large number of iterations will lead to an "overfitting" and move us away from configurations, where the OSD post-processing step guesses the logical error correctly. A similar observation was also reported in Ref.~\cite{higgott2023improved}.
Increasing the OSD order gives improved results at the cost of run time, which is why we limit the order to \(60\).

\begin{table}
    \caption{Parameters used for BP+OSD decoding. Here, \([[n,k,d]]\) refer to the code parameters}
    \label{tab:bp_params}
    \centering
    \begin{ruledtabular}
    \begin{tabular}{cc}
        Parameter & Value  \\
        \hline
        BP method & product sum \\
        maximum BP iterations & \(\lfloor\frac{d}{2}\rfloor\) \\
        osd order & \(\min(d^2,60)\)
    \end{tabular}
    \end{ruledtabular}
\end{table}

\begin{figure}
    \centering
    \includegraphics[width=\linewidth]{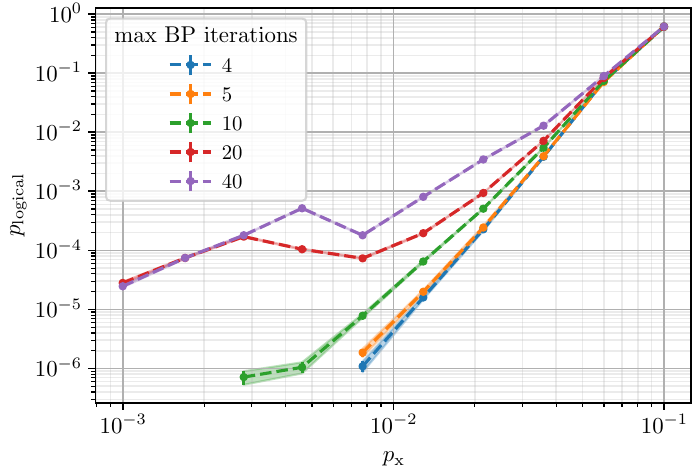}
    \caption{Logical error rate for the \([[369,9,9]]\)-code using BP+OSD with different maximum BP iterations. Using only \(4\) iterations gives the lowest logical error rate.}
    \label{plot:bposd_9_maxiter}
\end{figure}

\begin{figure}
    \centering
    \includegraphics[width=\linewidth]{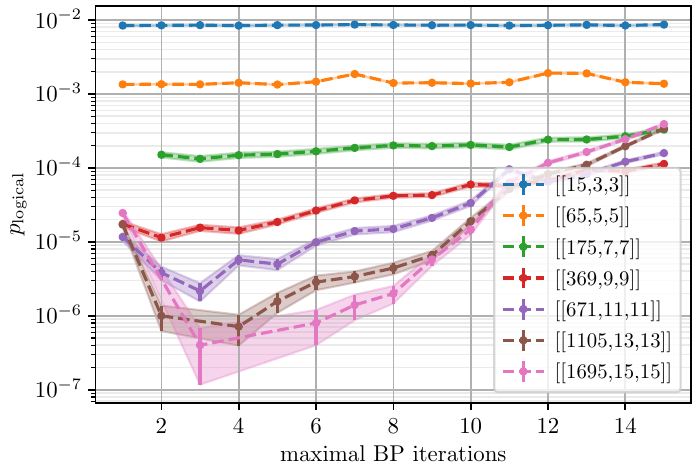}
    \caption{Logical error rates for the LCS-codes of family 2 (see main text Sec.~\ref{sec:codefamilies}) using BP+OSD with different maximum BP iterations at a fixed physical error rate \(p = 0.0129\). In particular for larger codes, the logical error rate is highly dependent on the maximum number of BP iterations before OSD post-processing.}
    \label{plot:bposd_maxiter}
\end{figure}

\section{Validity of BP+OSD decoding}\label{app:bpvalidity}
To verify the validity of BP+OSD decoding, we compare the most-likely error decoder to the BP+OSD decoder. As can be seen in Fig.~\ref{plot:mlevsbposd}, they perform very similarly for small qubit numbers, but for many qubits and low error rate, the decoding performance of BP+OSD decreases.

We compare the performance of BP+OSD with pymatching in Fig.~\ref{plot:pymvsbposd}. While it is typically observed that decoding surface codes with BP+OSD results in a lower threshold (cf. Ref.~\cite{roffe2020decoding}), we observe that BP+OSD achieves lower logical error rates, in particular for the phenomenological noise model. For this reason, we compare the results for LCS codes with results for surface codes decoded with BP+OSD in the main text.

\begin{figure}
    \centering
    \includegraphics[width=\linewidth]{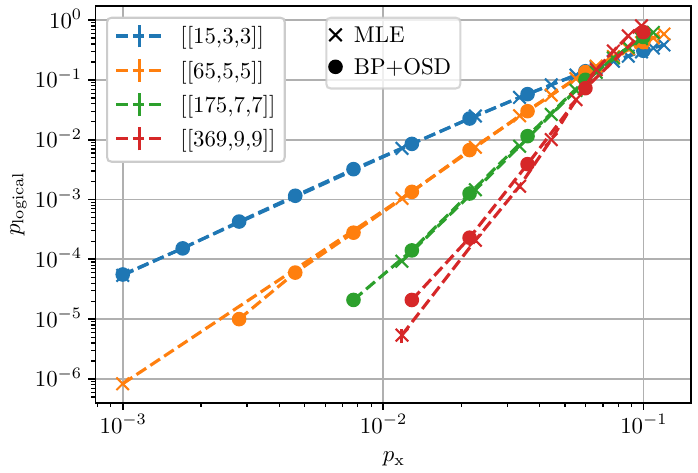}
    \caption{Logical error rate for LCS codes (family 2, see main text Sec.~\ref{sec:codefamilies}) using MLE and BP+OSD. In particular for low qubit numbers, BP+OSD comes close to MLE decoding.}
    \label{plot:mlevsbposd}
\end{figure}

\begin{figure}
    \centering
    \includegraphics[width=\linewidth]{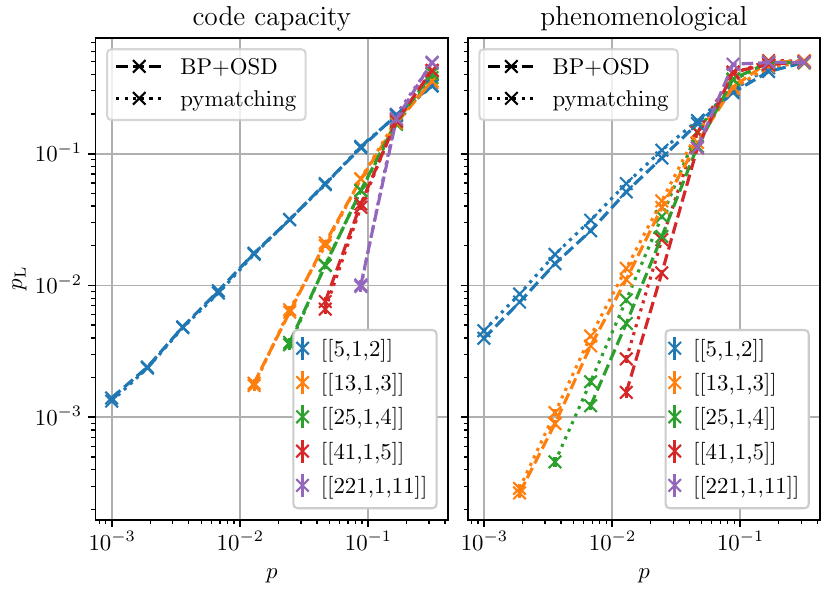}
    \caption{Logical error rates for surface codes using pymatching and BP+OSD with a code capacity (left) and phenomenological (right) noise model. BP+OSD achieves a comparable or lower logical error rate.}
    \label{plot:pymvsbposd}
\end{figure}

\section{Sampling Methods}
The implementation details of the sampling for code capacity noise and phenomenological noise are shown in Alg.~\ref{alg:cc_sampling} and Alg.~\ref{alg:phn_sampling} respectively.
\begin{algorithm}
        \SetAlgoLined
        \DontPrintSemicolon
        \BlankLine
        \KwIn{Quantum Channel \(\mc{E}\), physical error probabilities \(\vb{p}_{\text{phys}} = (p_x, p_y, p_z)\), number of samples \(n\), Quantum Code \(\mc{Q}\)}
        \KwOut{Number of successful runs \(n_{\text{success}}\)}
        \(n_{\text{success}} = 0 \) \;
        \For{\(i = 0\), \(i < n\), \(i = i+1\)}{
        \(E = \mc{E}(0, \vb{p})\) \;
        \(\vb{s} = \sigma(E)\) \;
        \(E^\star = \mathrm{Dec}(\vb{s}, \vb{p}, \dots)\) \;
        \If{\(\sigma(E E^\star) = 0 \; \& \; \langle E E^\star, L \rangle = 0 \quad \forall L \in \mc{L}(\mc{Q})\)}
        {\(n_{\text{success}} = n_{\text{success}} + 1 \)} 
        }
        \KwRet{\(n_{\text{success}}\)}

        \caption{Code capacity sampling of quantum codes}
        \label{alg:cc_sampling}
\end{algorithm}

\begin{algorithm}
        \SetAlgoLined
        \DontPrintSemicolon
        \BlankLine
        \KwIn{Quantum Channel \(\mc{E}\), physical qubit error probabilities \(\vb{p}_{\text{phys}} = (p_x, p_y, p_z)\), Classical Channel \(\mc{B}\), physical syndrome error probability \(q\), number of samples \(n\), Quantum Code \(\mc{Q}\)}
        \KwOut{Number of successful runs \(n_{\text{success}}\)}
        \(n_{\text{success}} = 0 \) \;
        \For{\(i = 0\), \(i < n\), \(i = i+1\)}{
        \(E = 0\) \;
        \(\vb{s} = (0)^{((t+1) \times n_c)}\) \;
        \For{\(t = 0\), \(t , d(\mc{Q})\), \(t = t+1\)}
        {
            \(E = \mc{E}(E, \vb{p})\) \;
            \(\vb{s}_{t+1} = \mc{B}(\sigma(E))\) \;
        }
        \(E^\star = \mathrm{Dec_1}(\vb{s}, \vb{p}, q, \dots)\) \;
        \(\vb{s}^\star = \sigma(E E^\star)\)  \;
        \(E^{\star \star} = \mathrm{Dec_2}(\vb{s}^\star, \vb{p}, \dots)\)  \;
        \If{\(\sigma(E E^\star E^{\star \star}) = 0 \; \& \; \langle E E^\star E^{\star \star}, L \rangle = 0 \quad \forall L \in \mc{L}(\mc{Q})\)}
        {\(n_{\text{success}} = n_{\text{success}} + 1 \)}
        }
        \KwRet{\(n_{\text{success}}\)}

        \caption{Phenomenological noise sampling of quantum codes}
        \label{alg:phn_sampling}
\end{algorithm}

\clearpage
\bibliography{lcs_codes}

\end{document}
%
% ****** End of file lcs_codes.tex ******

%